\documentclass[%
 reprint,
 preprintnumbers,
 nofootinbib,
 amsmath,amssymb,
 aps,
]{revtex4-2}

\usepackage{graphicx}
\usepackage{dcolumn}
\usepackage{bm}
\usepackage{mathtools}
\usepackage{mathrsfs}
\usepackage{amssymb}
\usepackage{hyperref}
\usepackage[usenames,dvipsnames]{xcolor}
\hypersetup{
    colorlinks=true,
    citecolor=NavyBlue,
    linkcolor=BrickRed,
    urlcolor=CadetBlue,
}
\newcommand{\cE}{\mathcal{E}}
\newcommand{\cF}{\mathcal{F}}
\newcommand{\cG}{\mathcal{G}}
\newcommand{\cH}{\mathcal{H}}
\newcommand{\cL}{\mathcal{L}}
\newcommand{\cP}{\mathcal{P}}
\newcommand{\cQ}{\mathcal{Q}}
\newcommand{\cR}{\mathcal{R}}
\newcommand{\rh}{\mathrm{h}}
\newcommand{\rL}{\mathrm{L}}
\newcommand{\rR}{\mathrm{R}}
\newcommand{\bA}{\mathbf{A}}
\newcommand{\bM}{\mathbf{M}}
\newcommand{\bR}{\mathbf{R}}
\newcommand{\ba}{\mathbf{a}}
\newcommand{\defi}{\coloneqq}
\newcommand{\til}{\widetilde}

\newcommand{\pd}{\partial}
\newcommand{\Mpl}{M_{\mathrm{Pl}}}

\allowdisplaybreaks[4]

\begin{document}

\preprint{KOBE-COSMO-21-16}

\title{Quasinormal modes of charged black holes with corrections \\ from nonlinear electrodynamics}

\author{Kimihiro Nomura}
 \email{knomura@stu.kobe-u.ac.jp}
 
\author{Daisuke Yoshida}%
 \email{dyoshida@hawk.kobe-u.ac.jp}
\affiliation{%
Department of Physics, Kobe University, Kobe 657-8501, Japan
}%

\date{\today}

\begin{abstract}
We study quasinormal modes related to gravitational and electromagnetic perturbations of spherically symmetric charged black holes in nonlinear electrodynamics. Beyond the linear Maxwell electrodynamics, we consider a class of Lagrangian with higher-order corrections written by the electromagnetic field strength and its Hodge dual with arbitrary coefficients, and we parametrize the corrections for quasinormal frequencies in terms of the coefficients. It is confirmed that the isospectrality of quasinormal modes under parity is generally violated due to nonlinear electrodynamics. As applications, the corrections for quasinormal frequencies in Euler--Heisenberg and Born--Infeld electrodynamics are calculated, then it is clarified that the nonlinear effects act to lengthen the oscillation period and enhance the damping rate of the quasinormal modes.
\end{abstract}

\maketitle


\section{Introduction}
\label{sec:Introduction}

General relativity, which relates gravity to the geometry of spacetime, has revealed a spacetime structure from which nothing can escape, called a black hole. It is now known that black holes are abundantly present in our universe. Black holes are fascinating objects from an observational point of view since they can be a source of gravitational waves, for which observations have been greatly advanced in recent years. Gravitational waves are produced, for example, by coalescence of a binary and infalling of matters. An important feature of such waves propagating from a black hole is that they undergo characteristic damping oscillations, called quasinormal modes, as it settles into a stationary state (for reviews, see Refs.\,\cite{Kokkotas:1999bd, Nollert:1999ji, Berti:2009kk, Konoplya:2011qq}). General relativity predicts that the frequencies of the quasinormal modes are uniquely determined by a few black hole parameters, such as the mass, charge, and spin. Therefore, quasinormal modes are important observables for estimating parameters of black holes in our universe, and for testing the validity of general relativity \cite{Berti:2015itd}.

If once a deviation of quasinormal modes from the prediction by general relativity is detected, it suggests the need to take into account some new effects. To explore physics beyond the standard theory, it is important to investigate quasinormal modes of black holes taking in new effects. In fact, there have been many attempts to extend the theory of gravity, and the quasinormal modes of vacuum black hole solutions have been studied in extensions of general relativity \cite{Cardoso:2009pk, Molina:2010fb, Blazquez-Salcedo:2016enn, Blazquez-Salcedo:2017txk, Tattersall:2017erk, Cardoso:2018ptl, Tattersall:2018nve, Fontana:2018fof, Cardoso:2019mqo, McManus:2019ulj, Abdalla:2019irr, Destounis:2019omd, deRham:2020ejn, Konoplya:2020bxa, Cano:2021myl, Moura:2021eln, Moura:2021nuh, Pierini:2021jxd, Wagle:2021tam, Srivastava:2021imr}. On the other hand, for electrically or magnetically charged black holes, the Einstein--Maxwell theory is usually considered to be the standard theory and leads to the so-called Reissner--Nordstr\"om solution, for which the quasinormal frequencies are calculated in Refs.\,\cite{gunter1980study, Kokkotas:1988fm, Leaver:1990zz, andersson1993normal}. However, the quasinormal modes may be subject to corrections due to some effects of electrodynamics beyond the Maxwell theory, which actually exist at least when one considers quantum electrodynamics as we will see below. With this in mind, this paper is devoted to studying the quasinormal modes of charged black holes in a framework beyond the standard Einstein--Maxwell theory.

Until today, there is no evidence for the existence of charged black holes in our universe although there have been attempts to search them e.g.~by analyzing LIGO-Virgo data \cite{Carullo:2021oxn}. However, the possibility of their existence has often been discussed. For example, in Refs.\,\cite{Zajacek:2018ycb, Zajacek:2019kla}, it is pointed out that a black hole should have a slight electric charge by assuming a balance between the number of protons and electrons around the black hole. 
Furthermore, a rotating black hole in a uniform magnetic field can have an electric charge proportional to its spin, as known as the Wald solution \cite{Wald:1974np}. 
On the other hand, magnetically charged black holes may be produced in the early universe, and they are more likely to retain the magnetic charge avoiding the neutralization by ordinary matter accreting on them unlike electrically charged counterparts \cite{Maldacena:2020skw}. Therefore, electrically or magnetically charged black holes are not merely objects of mathematical interest, and it is worth studying the quasinormal modes of them to correctly capture the parameters of realistic black holes from observations.

The effects of electrodynamics beyond the Maxwell theory can appear, for example, when one considers quantum vacuum polarization. An example of the effective Lagrangian taking in the electron one-loop corrections is known as the Euler--Heisenberg Lagrangian \cite{Heisenberg:1936nmg}. It includes higher-order terms of the electromagnetic field in addition to the Maxwell Lagrangian, such as $(F^{\mu\nu}F_{\mu\nu})^2$ and $(F^{\mu\nu}\widetilde{F}_{\mu\nu})^2$ with suppressions proportional to $m_e^{-4}$, where $F_{\mu\nu}$ is the electromagnetic field strength tensor, $\til{F}_{\mu\nu}$ is its Hodge dual, and $m_e$ is the electron mass. More generally, the effective Lagrangian of electrodynamics incorporating some high energy effects is considered to be given by a general function of $F_{\mu\nu}$ and $\til{F}_{\mu\nu}$. Such a framework is collectively referred to as nonlinear electrodynamics. An old example of nonlinear electrodynamics is Born--Infeld theory, which was proposed in the 1930s to remove the divergence of the electron's self-energy within classical electrodynamics \cite{Born:1934gh}. 

Then, it has been motivated to investigate black hole solutions in general relativity coupled to such nonlinear electrodynamics as in Refs.\,\cite{Peres:1961zz, deOliveira:1994in}. In fact, black hole solutions are actively studied in the Euler--Heisenberg theory \cite{Yajima:2000kw,Ruffini:2013hia, Allahyari:2019jqz, Amaro:2020xro, Breton:2021mju} and in the Born--Infeld theory \cite{Hoffmann:1937noa, Pellicer:1969cf, Demianski:1986wx, Breton:2003tk, Breton:2007bza, Kruglov:2017mpj}. 
A more interesting aspect of general relativity with nonlinear electrodynamics is that one can construct black holes without spacetime singularity, such as the Bardeen black hole \cite{AyonBeato:2000zs}. So far, various regular black holes have been obtained as solutions in specific theories of nonlinear electrodynamics \cite{AyonBeato:1998ub, AyonBeato:1999ec,AyonBeato:1999rg, Bronnikov:2000vy, AyonBeato:2004ih, Fernando:2016ksb, Fan:2016hvf, Chinaglia:2017uqd, Bronnikov:2017sgg, Rodrigues:2018bdc, Ali:2018boy, Poshteh:2020sgp, Villani:2021lmo, Mehdipour:2021ipf, Kruglov:2021mfy}.
For one of the models of such regular black holes, quasinormal modes are also analyzed in Ref.\,\cite{Chaverra:2016ttw}.
On the more general ground, one can study the properties of black holes in nonlinear electrodynamics with a Lagrangian given by a general function of the electromagnetic field. The perturbative stability of charged black holes has been investigated in theories with a Lagrangian being a general function of $F_{\mu\nu} F^{\mu\nu}$ in Ref.\,\cite{Moreno:2002gg}. Moreover, recently the perturbative stability of charged black holes is studied, and equations of motion for the perturbations on them are derived, in general nonlinear electrodynamics including $F_{\mu\nu} \til{F}^{\mu\nu}$ dependence in Refs.\,\cite{Nomura:2020tpc, Daghigh:2021psm}.

In this paper, we study quasinormal modes related to gravitational and electromagnetic perturbations of charged black holes in general relativity with nonlinear electrodynamics. In fact, quasinormal modes of black holes have already been analyzed for some specific theories of nonlinear electrodynamics e.g.~in Refs.\,\cite{Fernando:2004pc, Flachi:2012nv, Li:2014fka, Chaverra:2016ttw, Xi:2016qrg, Panotopoulos:2017hns, Wu:2018xza, Destounis:2018utr, Panotopoulos:2019tyg, Panotopoulos:2019qjk, Panotopoulos:2020zbj, Rincon:2021gwd, Okyay:2021nnh}. On the other hand, in this paper, we consider a general framework of nonlinear electrodynamics as an extension of Maxwell electrodynamics without specifying a certain theory, and aim to find the leading corrections for quasinormal modes due to the nonlinear effects. As a useful way for this purpose, we take the effective field theoretical approach, i.e., we consider a Lagrangian with higher-order terms of $F_{\mu\nu}$ and $\til{F}_{\mu\nu}$ in addition to the Einstein--Maxwell Lagrangian, and we calculate the corrections for quasinormal frequencies due to such terms. The higher-order terms possibly appear by integrating out some heavy degrees of freedom. But, from a purely bottom-up point of view, we take the coefficients of such terms to be arbitrary so that the framework of nonlinear electrodynamics including the Euler--Heisenberg and Born--Infeld theories can be treated inclusively. For simplicity, we will introduce only the terms of $(F_{\mu\nu} F^{\mu\nu})^2$ and $(F_{\mu\nu} \til{F}^{\mu\nu})^2$ with arbitrary coefficients, which are expected to give the leading corrections to the Maxwell electrodynamics in situations where the electromagnetic field strength is smaller than some cutoff. This kind of effective field theoretical approach in pure gravitational theory can be found in Ref.\,\cite{Cardoso:2018ptl}, where an extension of general relativity with higher-order terms of the Riemann tensor is considered, and quasinormal modes of the vacuum solution are studied.

This paper is organized as follows. In Sec.\,\ref{sec:Blackholes}, spherically symmetric black hole solutions in general relativity coupled to general nonlinear electrodynamics are reviewed, and master equations describing gravitational and electromagnetic perturbations on them are given. In Sec.\,\ref{sec:EFT}, we introduce the effective field theoretical approach in nonlinear electrodynamics. In Sec.\,\ref{sec:QNM}, we calculate the frequencies of quasinormal modes related to electromagnetic and gravitational perturbations of charged black holes in effective field theory of nonlinear electrodynamics. We independently use two calculation methods: numerical integration and the continued fractions method. Then, we apply the results to the Euler--Heisenberg and Born--Infeld theories. Section \ref{sec:Summary} is devoted to the summary and discussion. In Appendix \ref{app:duality}, we review the construction of magnetically or electrically charged black holes in nonlinear electrodynamics, and the dual relationship between them.

Throughout this paper, we take the following conventions. The metric signature is $(-,+,+,+)$, and the covariant anti-symmetric tensor is normalized as $\epsilon_{0123} = \sqrt{-g}$, where $g$ is the determinant of the metric.
We set $c = \hbar = \mu_0 = 1$, where $c$, $\hbar$, and $\mu_0$ are the speed of light, reduced Planck constant, and permeability in a vacuum, respectively.

\section{Black holes in nonlinear electrodynamics: Background and perturbations}
\label{sec:Blackholes}

In this section, we briefly review spherically symmetric black hole solutions in nonlinear electrodynamics coupled to general relativity, and write down master equations describing gravitational and electromagnetic perturbations on them. The descriptions here are almost based on the previous work by the authors \cite{Nomura:2020tpc}, and references therein.

\subsection{Black hole background}
\label{subsec:Blackholebackground}

We start with the action in general relativity with nonlinear electrodynamics,
\begin{align}
    S[g_{\mu\nu}, A_\mu] = \int d^4 x \sqrt{-g} \left[ \frac{1}{16 \pi G} R- \cL(\cF, \cG) \right]\,,
    \label{eqaction}
\end{align}
where $G$ is the gravitational constant, $R$ is the Ricci scalar calculated from the metric $g_{\mu\nu}$, and $\cL(\cF, \cG)$ is an arbitrary function of $\cF$ and $\cG$ defined by 
\begin{align}
    \cF &\defi \frac{1}{4} F_{\mu\nu} F^{\mu\nu}\,,\\
    \cG &\defi \frac{1}{4} F_{\mu\nu} \til{F}^{\mu\nu} = \frac{1}{8} \epsilon_{\mu\nu\rho\sigma} F^{\mu\nu} F^{\rho\sigma}\,.
\end{align}
Here, $F_{\mu\nu} = \pd_\mu A_\nu - \pd_\nu A_\mu$ is the field strength tensor of the electromagnetic (abelian gauge) field $A_\mu$, and $\til{F}_{\mu\nu} = (1/2) \epsilon_{\mu\nu\rho\sigma} F^{\rho\sigma}$ is the Hodge dual of $F_{\mu\nu}$.
The Lagrangian in terms of two scalars, $\cL(\cF, \cG)$, is generic one in nonlinear electrodynamics in the sense that arbitrary invariants constructed from $F_{\mu\nu}$ and $\til{F}_{\mu\nu}$ can be reduced to that form, as explained in Appendix A of Ref.\,\cite{Nomura:2020tpc}.

\subsubsection{Magnetic black holes}

Let us study spherically symmetric solutions with magnetic charge at first.
To consider asymptotically flat solutions for simplicity, we assume that the Lagrangian \eqref{eqaction} does not have a cosmological constant.
While here we simply show the results, the more detailed derivations can be found in Appendix \ref{app:magnetic} of the present paper, or Ref.\,\cite{Nomura:2020tpc}.
We take the spherically symmetric configuration of the magnetic field as
\begin{align}
    \frac{1}{2} F_{\mu\nu} dx^\mu \wedge dx^\nu = q \sin \theta d\theta \wedge d\phi\,,
    \label{eqmagbg}
\end{align}
where $q$ is a constant corresponding to the magnetic charge.
Note that on this magnetic configuration we have
\begin{align}
    \cF = \frac{q^2}{2 r^4}\,, \quad \cG = 0\,.
    \label{eqbg}
\end{align}
Then, we obtain the spherically symmetric solution as 
\begin{align}
    g_{\mu\nu} dx^\mu dx^\nu = - f(r) dt^2 + \frac{1}{f(r)} dr^2 + r^2 (d\theta^2 + \sin^2 \theta  d\phi^2) \,,
    \label{eqmetric1}
\end{align}
where 
\begin{align}
    f(r) &= 1 - \frac{2GM}{r} - \frac{8 \pi G}{r} \int^r dr' \, {r'}^2 \cL \left( \frac{q^2}{2{r'}^4} , 0 \right)\,.
    \label{eqmetric2}
\end{align}
Here, a constant $M$ corresponds to the mass of the gravitating object.
When the function $f(r)$ vanishes at some point(s),  the metric describes the magnetic black hole.
We write the position of the outer event horizon as $r_\rh$, i.e., $r_\rh$ is the largest radial coordinate such that 
\begin{align}
    f(r_\rh) = 0\,.
\end{align}

\subsubsection{Electric black holes}

To find electrically charged black hole solutions, it is convenient to move from the original framework in terms of $\cL(\cF, \cG)$ to an alternative framework by the Legendre transformation; the explanation is given in Appendix \ref{app:electric}.
One important thing we should emphasize here is that an electric black hole solution in some theory of nonlinear electrodynamics can be translated into a magnetic black hole solution in another theory via a kind of duality, and vice versa, as mentioned in Appendix \ref{app:FPduality}.
For this reason, we thoroughly work on magnetic black holes below, and the results can be rendered into those on the electric background in an appropriate way.
As a concrete example, we study the duality for an effective Lagrangian up to the quadratic order of $\cF$ and $\cG$ in Appendix \ref{app:FPdualityEFT}, and the results will be mentioned in the next section.

\subsection{Master equations for black hole perturbations}
\label{subsec:perturbations}

Here we review master equations which describe linear perturbations of the metric and electromagnetic field on the magnetic black hole background continuing to follow Ref.\,\cite{Nomura:2020tpc}.
On the spherically symmetric background, the linear perturbations can be expanded on the basis of tensor spherical harmonics.
Moreover, on the magnetic background, the linear perturbations are separated into two systems according to parity;
on the one hand, the odd-parity metric and the even-parity electromagnetic perturbations are coupled, which we call the type I;
on the other hand, the even-parity metric and the odd-parity electromagnetic perturbations are coupled, which we call the type II.
In each type of system, the master equation is reduced to a Schr\"odinger-like equation in a matrix form with an effective potential, which we write as
\begin{align}
    \left( \frac{d^2}{dr^{*2}} + \omega^2 - f V_{\mathrm{I, II}} \right) 
    \begin{pmatrix}
        \cR_{\mathrm{I, II}}\\
        \cE_{\mathrm{I, II}}
    \end{pmatrix}
    = 0\,
    \label{eqmaster}
\end{align}
with a Fourier mode $\omega$.
Here, $r^*$ is the tortoise coordinate defined by
\begin{align}
    r^* = \int^r \frac{dr'}{f(r')}\,, 
\end{align}
then $r = r_\rh$ corresponds to $r^* = -\infty$. We write gauge invariant perturbations of the metric and electromagnetic field as $\cR_{\mathrm{I, II}}$ and $\cE_{\mathrm{I, II}}$, respectively, where the subscripts I and II represent that the quantities belong to each type.

The effective potentials $fV_{\mathrm{I},\mathrm{II}}$ in Eq.\,\eqref{eqmaster} are given by $2 \times 2$ symmetric matrices.
For the effective potential of the type I,
\begin{align}
    fV_\mathrm{I} = f
    \begin{pmatrix}
        V_{\mathrm{I, 11}} & V_{\mathrm{I, 12}}\\
        V_{\mathrm{I, 21}} & V_{\mathrm{I, 22}}
    \end{pmatrix}\,,
\end{align}
the components are given by 
\begin{align}
    V_{\mathrm{I, 11}} &= \frac{l(l+1) + 3(f-1)}{r^2} + 8 \pi G \cL \,,
    \label{eqVI11}\\
    V_{\mathrm{I, 12}} = V_{\mathrm{I, 21}} &= \frac{\sqrt{16 \pi G q^2 \cL_\cF (l+2)(l-1)}}{r^3} \,,
    \label{eqVI12}\\
    V_{\mathrm{I, 22}} &= \frac{l(l+1)}{r^2} \frac{1}{1 - q^2 \cL_{\cG\cG} / (r^4 \cL_\cF)} + \frac{16 \pi G q^2 \cL_\cF}{r^4} \notag \\
    &\quad - \left( 8 \pi G \cL + \frac{6f-1}{r^2} \right) \frac{q^2 \cL_{\cF\cF}}{r^4 \cL_\cF}  \notag \\
    &\quad + \frac{3fq^4 \cL_{\cF\cF}^2}{r^{10} \cL_\cF^2 } - \frac{2fq^4 \cL_{\cF\cF\cF}}{r^{10} \cL_\cF}\,.
    \label{eqVI22}
\end{align}
Here, $\cL$ is interpreted as the background value obtained by inserting Eq.\,\eqref{eqbg}, i.e., $\cL = \cL(\cF = q^2/(2r^4), \cG = 0)$. 
Similarly, we write as $\cL_\cF = (\pd \cL/ \pd \cF)(q^2/(2r^4), 0)$, $\cL_{\cF\cF} = (\pd^2 \cL/ \pd \cF^2)(q^2/(2r^4), 0)$, etc.
The integer $l$ labels the multipole of the perturbations.
We are interested in the modes of $l \geq 2$ where both the metric and electromagnetic perturbations are dynamical.
Note that we assume that the Lagrangian satisfies $\cL_\cF > 0$ to avoid the ghost instability, and $\cL_{\cG} = 0$ for the existence of the black hole solution \cite{Nomura:2020tpc}.

The effective potential of the type II is given by 
\begin{align}
    fV_\mathrm{II} =f
    \begin{pmatrix}
        V_{\mathrm{II, 11}} & V_{\mathrm{II, 12}}\\
        V_{\mathrm{II, 21}} & V_{\mathrm{II, 22}}
    \end{pmatrix}
\end{align}
with
\begin{widetext}
\begin{align}
    V_{\mathrm{II, 11}} &=\frac{\zeta(r)}{r^{2}} - \frac{2(2\lambda - f + 1)}{r^{2}} + \frac{8 \lambda (\lambda - f + 1)}{r^{2} \zeta(r)} + \frac{8 \lambda^{2} f}{r^{2} \zeta^{2}(r)}+ \frac{64 \pi G f q^{2} \lambda \cL_\cF}{r^{4} \zeta^{2}(r)} \,,
    \label{eqVII11}\\
    V_{\mathrm{II, 12}} = V_{\mathrm{II, 21}} &= 
    \sqrt{32 \pi G q^2 \lambda \cL_\cF}
    \left[ - \frac{1}{r^{3}} \left( 1 - \frac{2(2\lambda - f + 2)}{\zeta(r)} - \frac{4 \lambda f}{\zeta^{2}(r)} \right) + \frac{32 \pi G f q^{2} \cL_{\cF}}{r^{5} \zeta^{2}(r)} + \frac{2 f q^{2} \cL_{\cF\cF}}{r^{7} \zeta(r) \cL_{\cF}} \right]\,,
    \label{eqVII12}\\
    V_{\mathrm{II, 22}} &= \frac{2(\lambda + 1)}{r^{2}}- \frac{16 \pi G q^{2} \cL_{\cF}}{r^{4}}
    \left( 1 - \frac{4(\lambda+1)}{\zeta(r)} - \frac{4 \lambda f}{\zeta^{2}(r)} \right) 
    + \frac{512 \pi^2 G^2 fq^{4} \cL_{\cF}^{2} }{r^{6} \zeta^{2}(r)} 
    \notag \\
    &\quad - \frac{q^{2} \cL_{\mathcal{FF}}}{r^{6} \cL_{\cF}} \left[ \zeta(r) - 3f - 4(\lambda + 1) \right] + \frac{64 \pi G f q^{4} \cL_{\mathcal{FF}}}{r^{8} \zeta(r)} + \frac{2 f q^{4} \cL_{\mathcal{FFF}}}{r^{10} \cL_{\cF}} - \frac{f q^{4} \cL_{\mathcal{FF}}^{2}}{r^{10} \cL_{\cF}^{2}} \,,
    \label{eqVII22}
\end{align}
\end{widetext}
where we defined
\begin{align}
    \lambda &\defi \frac{1}{2}(l+2)(l-1) \,,\\
    \zeta (r) &\defi - 3f + 2\lambda  +3 - 8 \pi G r^{2}\cL \,.
\end{align}


\section{Effective Field Theory in nonlinear electrodynamics}
\label{sec:EFT}

The master equations reviewed in the previous section can be used to study linear perturbations on black holes for any nonlinear electrodynamics, but the analysis is complicated in general. However, to find the leading corrections from nonlinear electrodynamics, it is reasonable to consider the effective Lagrangian which consists of operators at the quadratic order in $\cF$ and $\cG$ in addition to the Maxwell Lagrangian.
Thus, we consider the action
\begin{align}
    S[g_{\mu\nu}, A_\mu] = \int d^4 x \sqrt{-g} \left[ \frac{1}{16 \pi G} R- \cL(\cF, \cG) \right]
\end{align}
with
\begin{align}
    \cL(\cF, \cG) = \cF - \alpha \cF^2 - \beta \cG^2\,,
    \label{eqeff}
\end{align}
where $\alpha$ and $\beta$ are parameters with the dimension $-4$.
For example, an effective Lagrangian of this form is obtained from the electron one-loop corrections, which is known as the Euler--Heisenberg Lagrangian and given by
\begin{align}
    \cL(\cF, \cG) = \cF - \frac{2}{45 m_e^4} \left( \frac{e^2}{4 \pi} \right)^2 (4 \cF^2 + 7 \cG^2)\,,
\end{align}
where $m_e$ is the mass of the electron and $e^2/(4\pi)$ is the fine structure constant.
In this paper, we intensively study the class of effective field theory of electrodynamics \eqref{eqeff} including the Euler--Heisenberg theory, and we aim to parametrize the corrections to the Maxwell theory in terms of $\alpha$ and $\beta$.

The magnetic black hole geometry in the effective theory can be easily obtained via Eq.\,\eqref{eqmetric2}.
For the Lagrangian \eqref{eqeff}, we have
\begin{align}
    \cL\left( \frac{q^2}{2r^4},0 \right) = \frac{q^2}{2r^4} - \frac{\alpha q^4}{4 r^8}\,,
\end{align}
so we obtain the function $f(r)$ in the metric \eqref{eqmetric1} as
\begin{align}
    f(r) &= 1 - \frac{2GM}{r} + \frac{4 \pi G q^2}{r^2} - \frac{2 \pi G \alpha q^4}{5 r^6}
    \notag \\
    &= 1 - \frac{2GM}{r} + \frac{Q^2}{r^2} - \frac{\bar{\alpha} Q^2 (GM)^4}{10 r^6}\,,
    \label{eqEFTbg}
\end{align}
where we defined 
\begin{align}
    Q^2 &\defi 4 \pi G q^2\,,
\end{align}
and introduced a dimensionless parameter 
\begin{align}
    \bar{\alpha} \defi \frac{\alpha q^2}{(GM)^4}.
    \label{eqpatalpha}
\end{align}
For later convenience, similarly we define
\begin{align}
    \bar{\beta} \defi \frac{\beta q^2}{(GM)^4}.
    \label{eqpatbeta}
\end{align}

We should note the validity of the effective Lagrangian \eqref{eqeff}. 
Since the background value of $\cF$ is given by Eq.\,\eqref{eqbg}, we have
\begin{align}
    \frac{\alpha \cF^2}{\cF} = \frac{\alpha q^2}{2 r^4}\,.
\end{align}
To justify the truncation of higher-order terms in the Lagrangian, this ratio should be suppressed outside the horizon $r_\rh \sim GM$, hence it is required that
\begin{align}
    \bar{\alpha} \ll 1\,.
\end{align}
Similarly, it is natural to expect
\begin{align}
    \bar{\beta} \ll 1\,.
\end{align}
Thus, the dimensionless quantities $\bar{\alpha}$ and $\bar{\beta}$ serve as the perturbation parameters.
Also note that we do not include non-minimal couplings between electromagnetic field and spacetime curvature.
In fact, up to four-derivative, such couplings can be listed as
\begin{align}
    &\gamma_1 R F_{\mu\nu} F^{\mu\nu}\,, ~
    \gamma_2 R_{\mu\nu} F^{\mu\rho} {F^\nu}_\rho \,, ~
    \gamma_3 R_{\mu\nu\rho\sigma} F^{\mu\nu} F^{\rho\sigma},
    \label{eqcoupling}
\end{align}
where $\gamma_{1,2,3}$ are parameters with dimension $-2$.
For example, taking account of electron one-loop corrections on curved spacetime yields $\gamma_{1,2,3} \sim (10^{-3} \text{--} 10^{-2}) \times e^2 / (4 \pi m_e^2)$ \cite{Drummond:1979pp}.
The contribution from $\gamma_1 R F_{\mu\nu} F^{\mu\nu}$ would be generally weaker than that from $\alpha \cF^2$ since $R$ vanishes in the Reissner--Nordstr\"om case.
Similarly, we can estimate the ratio of contributions from the last two operators in Eq.\,\eqref{eqcoupling} to $\alpha \cF^2$ on charged black holes as
\begin{align}
    \frac{\gamma_2 R_{\mu\nu} F^{\mu\rho} {F^\nu}_\rho}{\alpha \cF^2} &\sim 32 \pi \frac{ G \gamma_2}{\alpha}, 
    \\
    \frac{\gamma_3 R_{\mu\nu\rho\sigma} F^{\mu\nu} F^{\rho\sigma}}{\alpha \cF^2} &\sim - 64 \pi \frac{G \gamma_3}{\alpha} 
    + 128 \pi \left(\frac{GM}{Q} \right)^2 \frac{G \gamma_3}{\alpha},
\end{align}
where each quantity is evaluated at $r_\rh \sim GM$ since we are interested in dynamics around the horizon.
It is natural to consider the ratio $G\gamma_{1,2,3}/\alpha$ to be generally tiny in light of the example of the electron one-loop corrections, in which $G\gamma_{1,2,3}/\alpha \sim 10^{-1} Gm_e^2 (4\pi/e^2) \sim 10^{-44}$.
Therefore, it is expected that all three terms in Eq.\,\eqref{eqcoupling} are suppressed enough compared to Eq.\,\eqref{eqeff}, except for a case that the black hole is nearly free of charge.
Furthermore, higher-derivative terms such as $\gamma_4 \nabla_{\mu} F^{\mu\nu} \nabla_{\rho} {F^\rho}_{\nu}$ with a dimension $-2$ parameter $\gamma_4$ are also possible in the Lagrangian, but they are reduced to higher-order operators in perturbation parameters by means of equations of motion.
Therefore, from here we concentrate on the effective Lagrangian \eqref{eqeff} as giving the leading corrections, for simplicity.

The position of the event horizon, $r_\rh$, can be found perturbatively with respect to $\bar{\alpha}$.
In the linear order of $\bar{\alpha}$, we have
\begin{align}
    r_\rh = r_0 + \bar{\alpha} r_1 + \mathcal{O}(\bar{\alpha}^2)
    \label{eqrhexp}
\end{align}
where 
\begin{align}
    r_0 &= GM + \sqrt{(GM)^2 - Q^2}\,,
    \notag \\
    \bar{\alpha} r_1 &= \bar{\alpha}\frac{Q^2 (GM)^4}{20 r_0^3 (r_0 GM - Q^2)}\,.
\end{align}

The effective potentials in the master equations are immediately obtained by plugging
\begin{align}
    \cL &= \frac{q^2}{2r^4} - \frac{\alpha q^4}{4 r^8} = \frac{Q^2}{8 \pi G r^4} - \frac{\bar{\alpha} Q^2 (GM)^4}{16 \pi G r^8} \,,
    \label{eqeffpot1}\\
    \cL_\cF &= 1 - \frac{\alpha q^2}{r^4} = 1- \frac{\bar{\alpha} (GM)^4}{r^4}\,,\\
    \cL_{\cF\cF} &= -2\alpha = - \frac{8 \pi G \bar{\alpha} (GM)^4}{Q^2}\,,\\
    \cL_{\cG\cG} &= - 2\beta = - \frac{8 \pi G \bar{\beta} (GM)^4}{Q^2}\,,\\
    \cL_{\cF\cF\cF} &= 0\,,
    \label{eqeffpot5}
\end{align}
into Eqs.\,\eqref{eqVI11}--\eqref{eqVI22} for the type I, and into Eqs.\,\eqref{eqVII11}--\eqref{eqVII22} for the type II.
For convenience, here let us explicitly show them: for the type I,
\begin{widetext}
\begin{align}
    V_{\mathrm{I, 11}} &= \frac{l(l+1) + 3(f-1)}{r^2} + \frac{Q^2}{r^4} - \frac{\bar{\alpha} Q^2 (GM)^4}{2 r^8}\,,
    \label{eqEFTVI11}\\
    V_{\mathrm{I, 12}} = V_{\mathrm{I, 21}} &= \frac{2\sqrt{(l+2)(l-1) Q^2 [1 - \bar{\alpha} (GM)^4 / r^4]}}{r^3} \,,
    \label{eqEFTVI12}\\
    V_{\mathrm{I, 22}} &= \frac{l(l+1)}{r^2} \left( 1 + \frac{2 \bar{\beta} (GM)^4 / r^4}{1 - \bar{\alpha} (GM)^4 / r^4}  \right)^{-1} + \frac{4Q^2}{r^4} \left( 1 - \frac{\bar{\alpha}(GM)^4}{r^4} \right)
    \notag \\
    &\quad + \frac{2\bar{\alpha} (GM)^4}{r^6} \left( 1 - \frac{\bar{\alpha}(GM)^4}{r^4} \right)^{-1} \left(6f-1 + \frac{Q^2}{r^2} - \frac{\bar{\alpha}Q^2 (GM)^4}{2r^6} \right) + \frac{12 \bar{\alpha}^2 f (GM)^8}{r^{10}} \left( 1 - \frac{\bar{\alpha}(GM)^4}{r^4} \right)^{-2}\,,
    \label{eqEFTVI22}
\end{align}
and for the type II, 
\begin{align}
    V_{\mathrm{II, 11}} &=\frac{\zeta(r)}{r^{2}} - \frac{2(2\lambda - f + 1)}{r^{2}} + \frac{8 \lambda (\lambda - f + 1)}{r^{2} \zeta(r)} + \frac{8 \lambda^{2} f}{r^{2} \zeta^{2}(r)}+ \frac{16 \lambda f Q^2 }{r^{4} \zeta^{2}(r)} \left(1 - \frac{\bar{\alpha} (GM)^4}{r^4}\right) \,,
    \label{eqEFTVII11}\\
    V_{\mathrm{II, 12}} = V_{\mathrm{II, 21}} &= 
    \sqrt{8 \lambda Q^2 \left[1 - \frac{\bar{\alpha} (GM)^4}{r^4} \right]}
    \bigg\{ - \frac{1}{r^{3}} \left( 1 - \frac{2(2\lambda - f + 2)}{\zeta(r)} - \frac{4 \lambda f}{\zeta^{2}(r)} \right) + \frac{8 f Q^2}{r^{5} \zeta^{2}(r)} \left(1 - \frac{\bar{\alpha} (GM)^4}{r^4}\right)
     \notag \\
    &\qquad - \frac{4 \bar{\alpha} f (GM)^4}{r^7 \zeta(r)} \left(1 - \frac{\bar{\alpha} (GM)^4}{r^4}\right)^{-1} \bigg\}\,,
    \label{eqEFTVII12}\\
    V_{\mathrm{II, 22}} &= \frac{2(\lambda + 1)}{r^{2}}- \frac{4Q^2}{r^{4}} \left(1 - \frac{\bar{\alpha} (GM)^4}{r^4} \right) \left( 1 - \frac{4(\lambda+1)}{\zeta(r)} - \frac{4 \lambda f}{\zeta^{2}(r)} \right) 
    + \frac{32 f Q^4}{r^{6} \zeta^{2}(r)}  \left(1 - \frac{\bar{\alpha} (GM)^4}{r^4} \right)^2
    \notag \\
    &\quad + \frac{2 \bar{\alpha} (GM)^4\left( \zeta(r) - 3f - 4(\lambda + 1) \right)}{r^6} \left(1 - \frac{\bar{\alpha} (GM)^4}{r^4} \right)^{-1} 
    - \frac{32 \bar{\alpha} f Q^2 (GM)^4}{r^{8} \zeta(r)} 
    \notag \\
    &\quad - \frac{4 \bar{\alpha}^2 f (GM)^8}{r^{10}} \left(1 - \frac{\bar{\alpha} (GM)^4}{r^4} \right)^{-2} \,,
    \label{eqEFTVII22}
\end{align}
\end{widetext}
where 
\begin{align}
    \zeta(r) = -3 f + 2\lambda + 3 - \frac{Q^2}{r^2} + \frac{\bar{\alpha} Q^2 (GM)^4}{2r^6}\,,
\end{align}
and $f(r)$ is given by Eq.\,\eqref{eqEFTbg}.

\begin{figure*}
    \includegraphics[width=\hsize]{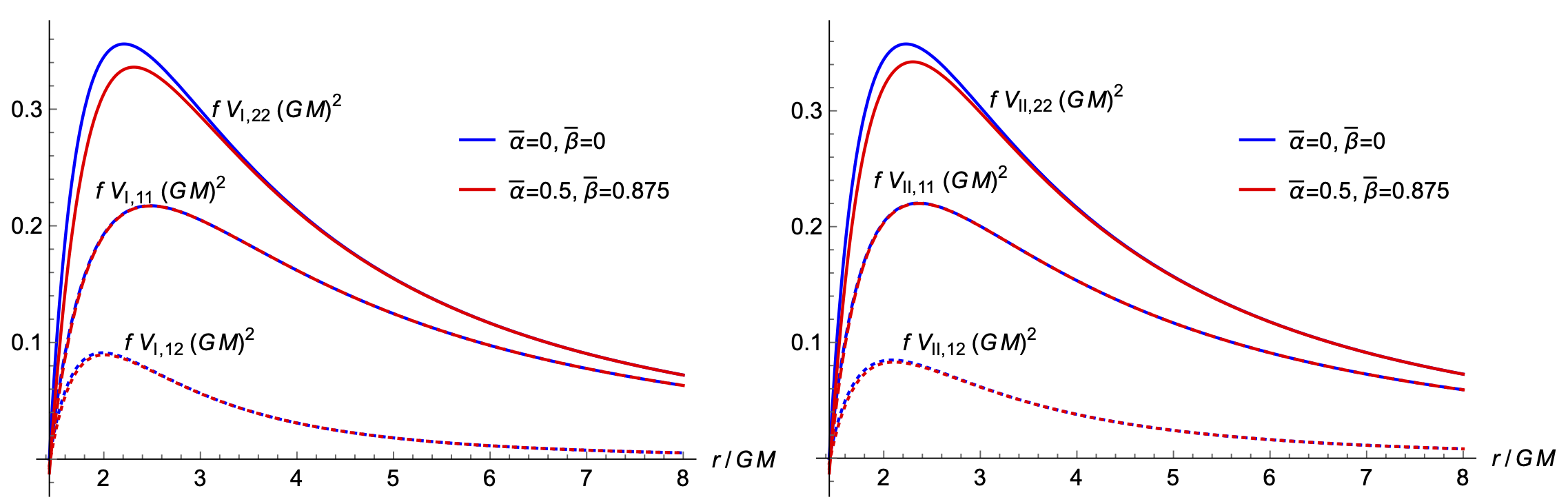}
    \caption{The components of the effective potentials in the master equations \eqref{eqmaster} for $l=2$ in the Maxwell electrodynamics are shown in blue, and those in the Euler--Heisenberg nonlinear electrodynamics are shown in red, as functions of $r$. (The quantities are normalized in terms of $GM$.) 
    The left panel is for the type I, and the right panel is for the type II. In each panel, and for each color, from top to bottom, the $fV_{22}$ component (solid curve), the $fV_{11}$ component (long dashed curve), and the $fV_{12}=fV_{21}$ component (short dashed curve) of the effective potential are shown.
    Here we take the charge-to-mass ratio of the black hole to be $Q/(GM) = 0.9$, and we choose the parameters in the Euler--Heisenberg nonlinear electrodynamics as $\bar{\alpha} = 0.5$ and $\bar{\beta} = 0.875$ to clarify the Euler--Heisenberg corrections. }
    \label{fig-pot}
\end{figure*}

In Fig.\,\ref{fig-pot}, the potential components for the $l=2$ mode on a magnetic black hole in the Maxwell electrodynamics (i.e., $\alpha = \beta = 0$) and in the Euler--Heisenberg nonlinear electrodynamics are shown as functions of $r$.
We can see that the Euler--Heisenberg correction given by $\alpha = (4/7) \beta > 0$ works to lower the height of the potential component $V_{22}$, while there are no noticeable corrections for $V_{11}$ and $V_{12}$.
The lowering of the height of $V_{22}$ due to the Euler--Heisenberg correction also appears for other modes of $l$.

So far we have focused on magnetically charged black holes, but electrically charged black holes in the effective theory \eqref{eqeff} are also given by the same geometry \eqref{eqEFTbg} up to the linear order in $\alpha$ and $\beta$, as explained in Appendix \ref{app:FPdualityEFT} of the present paper, and Ref.\,\cite{Ruffini:2013hia}.

\section{Quasinormal modes}
\label{sec:QNM}

\subsection{Definition}
\label{subsec:Definition}

Quasinormal modes are characteristic damping oscillation modes with complex eigenfrequencies of master equations, which are singled out by physically motivated boundary conditions.
In this section, we analyze quasinormal modes of the metric and electromagnetic perturbations on charged black holes in effective field theory of nonlinear electrodynamics defined by Eq.\,\eqref{eqeff}. 
Discrete complex eigenfrequencies corresponding to quasinormal modes are determined by boundary conditions at infinity and the event horizon as follows \cite{Chaverra:2016ttw}.
First, we let $f_\rR(\omega, r)$ be a $2 \times 2$ matrix-valued solution of the master equation which satisfies a boundary condition at infinity,
\begin{align}
    \lim_{r^* \to \infty}
    f_\rR (\omega, r) = e^{+i \omega r^*} \bm{1}\,,
    \label{eqbcinf}
\end{align}
where $\bm{1}$ is the $2 \times 2$ identity matrix.
On the other hand, we let $f_\rL(\omega, r)$ be a $2 \times 2$ matrix-valued solution of the master equation which satisfies a boundary condition at the event horizon,
\begin{align}
    \lim_{r^* \to -\infty}
    f_\rL (\omega, r) = e^{-i \omega r^*} \bm{1}\,.
    \label{eqbchor}
\end{align}
For $\mathrm{Im} (\omega) <0$, if there exist constant complex vectors $(a, b)^\mathrm{T}$ and $(c, d)^\mathrm{T}$ such that
\begin{align}
    \psi(\omega, r) = f_\rL (\omega, r) 
    \begin{pmatrix}
        a\\
        b
    \end{pmatrix}
    = f_\rR (\omega, r) 
    \begin{pmatrix}
        c\\
        d
    \end{pmatrix}\,,
    \label{eqpsi}
\end{align}
the $\omega$ is the quasinormal frequency and $e^{-i\omega t} \psi(\omega, r)$ represents the corresponding quasinormal mode.

As mentioned in Sec.\,\ref{subsec:perturbations}, the system of perturbations on black holes is separated into two types according to parity: the type I and II.
Each type has two families of quasinormal modes which we write as $Z_1$ and $Z_2$ with corresponding quasinormal frequencies $\omega_1$ and $\omega_2$, respectively.
In the Reissner--Nordstr\"om case, i.e., without nonlinear electrodynamics corrections, the effective potential matrices in the master equations can be diagonalized by constant matrices, thus $Z_1$ and $Z_2$ become decoupled and respectively obey single-component master equations \cite{Moncrief:1974ng, Moncrief:1975sb, Chaverra:2016ttw}.
On the other hand, in nonlinear electrodynamics, the decoupling does not work in general, so 
we utilize calculation methods valid for coupled systems as we will see below.

\subsection{Calculation methods}
\label{subsec:calculation}

\subsubsection{Numerical integration}
\label{subsubsec:numaric}

To compute quasinormal frequencies, we first use the numerical method developed in Refs.\,\cite{Chaverra:2015aya, Chaverra:2016ttw}.
While we here give the outline of methodology briefly, see the references for details. 
We first define an operator $T_{\rR, \omega}$ which acts on a $2 \times 2$ matrix-valued function $\xi$ as
\begin{align}
    (T_{\rR, \omega} \xi)(r) &= \bm{1} - \frac{1}{2i \omega} \int_{\gamma_{\theta_{1}}} dr' \left[ 1 - \exp \left( 2 i \omega \int_r^{r'} \frac{dr''}{f(r'')} \right) \right]
    \notag \\
    &\quad \times V(r') \xi(r') \,.
    \label{eqTR}
\end{align}
Here, $V(r)$ is the $2 \times 2$ potential matrix in the master equation. 
The integration path $\gamma_{\theta_{1}}$ is drawn in the complex $r$-plane and parametrized as
\begin{align}
    \gamma_{\theta_{1}} :~ r'(\lambda) = r + e^{i \theta_{1}} \lambda\,, 
\end{align}
where $\lambda$ runs from zero to positive infinity, and $\theta_{1}$ is an angle slightly larger than $-\pi/2$.
The integration from $r$ to $r'$ in the exponential is also performed along $\gamma_{\theta_1}$. 
Then, it is ensured that the integral converges for all $\omega$ with $\mathrm{Im}(\omega) < 0$ and $\mathrm{Re}(\omega) < 0$.
Note that the solution satisfying the boundary condition at infinity, Eq.\,\eqref{eqbcinf}, is obtained by iterating operation of $T_{\rR, \omega}$ as
\begin{align}
    f_\rR (\omega,r)
    = e^{+ i\omega r^*} \lim_{k \to \infty} ((T_{\rR, \omega})^k \bm{1})(r) \,.
    \label{eqfR}
\end{align}

Similarly, we define an operator $T_{\rL, \omega}$ which acts on a $2 \times 2$ matrix-valued function $\xi$ as
\begin{align}
    (T_{\rL, \omega} \xi)(r) &= \bm{1} + \frac{1}{2i \omega} \int_{\Gamma_{\theta_{2}}} dr' \left[ 1 - \exp \left( - 2i\omega \int_r^{r'} \frac{dr''}{f(r'')} \right) \right]
    \notag \\
    &\quad \times V(r') \xi(r') \,.
    \label{eqTL}
\end{align}
Here, the integration path $\Gamma_{\theta_{2}} (\lambda)$ is in the complex $r$-plane, and parametrized as
\begin{align}
    \Gamma_{\theta_{2}}:~ r'(\lambda) = r_\rh + (r - r_\rh) \exp(- e^{i\theta_{2}} \lambda)\,,
\end{align}
where $\lambda$ runs from zero to positive infinity, and the constant $\theta_{2}$ is slightly larger than $- \pi /2$.
The integration from $r$ to $r'$ in the exponential is also done along $\Gamma_{\theta_2}$.
The path $\Gamma_{\theta_{2}}$, which draws a counterclockwise spiral around the point $r_\rh$, is chosen so that the integral converges for all $\omega$ with $\mathrm{Im}(\omega) < 0$ and $\mathrm{Re}(\omega) < 0$.
Then, the solution satisfying the boundary condition at the event horizon, Eq.\,\eqref{eqbchor}, can be constructed by iterating operation of $T_{\rL, \omega}$ as
\begin{align}
    f_\rL (\omega, r)
    = e^{-i\omega r^*} \lim_{k \to \infty} ((T_{\rL, \omega})^k \bm{1})(r) \,.
    \label{eqfL}
\end{align}

The quasinormal frequencies are singled out by satisfaction of Eq.\,\eqref{eqpsi}, which implies that the Wronskian of the matrix-valued solutions $f_\rL(\omega, r)$ and $f_\rR(\omega, r)$ vanishes:
\begin{align}
    W(\omega)&= 
    \det\begin{pmatrix}
        f_\rL(\omega, r) & f_\rR(\omega, r) \\
        \pd_{r^*}f_\rL(\omega, r) & \pd_{r^*} f_\rR(\omega, r) 
    \end{pmatrix} 
    = 0 \,.
\end{align}
Therefore, we need to find $\omega$ such that $W(\omega) = 0$, where $f_\rR(\omega, r)$ and $f_\rL(\omega, r)$ are constructed by Eqs.\,\eqref{eqfR} and \eqref{eqfL}, respectively.

To find the solutions $f_\rL(\omega, r)$ and $f_\rR(\omega, r)$ numerically for a given $\omega$ with $\mathrm{Im}(\omega) < 0$ and $\mathrm{Re}(\omega) < 0$ according to the above procedure, we take the following treatments.
First, we fix as $\theta_{1} = -1.57$ and $\theta_{2} = -1.5$ following Refs.\,\cite{Chaverra:2015aya, Chaverra:2016ttw}.
We set a point for evaluating the Wronskian to be $r = 1.5 r_\rh$, at which we start the integrations.
We take the integration path $\gamma_{\theta_{1}}$, which should run infinitely in principle, to end at some point around $|r'| \sim 15$ (in the unit of $GM$), where the potential $V(r')$ is sufficiently small so that the integral does not result in the considerable value. 
Also, we take the end of the path $\Gamma_{\theta_{2}}$ at a point close enough to $r_\rh$.
The integrations along the paths are performed by trapezoidal approximation with about $5 \times 10^4$ steps.
We truncate the iterations after about 20 times, where the resulting values sufficiently converge.
In the end, we find quasinormal frequencies, i.e., zeros of the Wronskian, by Newton's method for complex $\omega$.

We are interested in quasinormal frequencies of black holes in effective nonlinear electrodynamics \eqref{eqeff}.
Thus, for magnetic black holes, the potentials $V(r)$ in Eqs.\,\eqref{eqTR} and \eqref{eqTL} are given by Eqs.\,\eqref{eqEFTVI11}--\eqref{eqEFTVI22} for the type I, and Eqs.\,\eqref{eqEFTVII11}--\eqref{eqEFTVII22} for the type II. We use these components for numerical calculations. As explained in the end of Sec.\,\ref{sec:EFT}, we can expect that the results are the same for electric black holes for sufficiently small $\bar{\alpha}$ and $\bar{\beta}$. Taking $GM$ as the unit, the system of our interest is characterized by three parameters: the black hole charge $Q$, and perturbation parameters $\bar{\alpha}$ and $\bar{\beta}$.
We compute quasinormal frequencies as varying these three parameters.

\subsubsection{Continued fractions method}
\label{subsubsec:fractions}

Apart from the numerical integration above, the continued fractions method is also known to be powerful to find quasinormal modes in a semi-analytical way, which is first applied to black hole perturbation theory by Leaver \cite{Leaver:1985ax}.
In a coupled system such as we are considering, the continued fractions method extended to a matrix-valued version is needed, some applications of which can be found in Refs.\,\cite{Pani:2013pma, Rosa:2011my, Pani:2012bp}.

First, let us assume that eigenfunctions of master equations \eqref{eqmaster} are given by the following series,
\begin{align}
    \begin{pmatrix}
        \cR(r)\\    
        \cE(r)
    \end{pmatrix}
     &= \left( \frac{r}{2GM} \right)^{2GMi\omega} e^{i\omega r} u^b \sum_{n=0}^\infty \begin{pmatrix}
        a^{(1)}_n\\    
        a^{(2)}_n
    \end{pmatrix}
    u^n,
    \label{eqansatz}
\end{align}
where $u \defi (r - r_\rh)/ r$, $b \defi - i \omega / f'(r_\rh)$, and $a^{(1,2)}_n$ are series coefficients.
The above ansatz takes account of the boundary conditions for quasinormal modes \eqref{eqbcinf} and \eqref{eqbchor}.
Indeed, at infinity, since $f(r) \simeq 1 - 2GM/r$, the tortoise coordinate is approximately given by
\begin{align}
    r^* \simeq r + 2GM \ln \left( \frac{r}{2GM} \right), \quad r \to \infty,
\end{align}
and thus the eigenfunctions corresponding to quasinormal modes should behave as
\begin{align}
    e^{i\omega r^*} \sim \left( \frac{r}{2GM} \right)^{2GMi\omega} e^{i\omega r},\quad r \to \infty.
\end{align}
On the other hand, near the horizon, since approximately $f(r) \simeq f'(r_\rh) (r-r_\rh)$, we have
\begin{align}
    r^* \simeq \frac{1}{f'(r_\rh)} \ln u, \quad r \to r_\rh,
\end{align}
and eigenfunctions should behave as
\begin{align}
    e^{-i\omega r^*} \sim u^{b}, \quad r \to r_\rh.
\end{align}
Then, one can see that the ansatz \eqref{eqansatz} reproduces these asymptotic behaviors appropriate to quasinormal modes.
Now, inserting the series \eqref{eqansatz} into master equations and extracting the homogeneous terms with respect to $u$, we can obtain a finite $k$-term recurrence relation for series coefficients as
\begin{align}
    \bA_{1,0} \ba_{1} + \bA_{2,0} \ba_{0} &= 0,
    \\
    \bA_{1,1} \ba_{2} + \bA_{2,1} \ba_{1} + \bA_{3,1} \ba_{0} &= 0,
    \\
    &\vdots
    \notag \\
    \bA_{1,k-3} \ba_{k-2} + \bA_{2,k-3} \ba_{k-3} &
    \notag \\
    + \bA_{3,k-3} \ba_{k-4} + \cdots + \bA_{k-1,k-3} \ba_{0} &= 0, 
    \\
    \bA_{1,n} \ba_{n+1} + \bA_{2,n} \ba_{n} + \bA_{3,n} \ba_{n-1} &
    \notag \\
     + \cdots + \bA_{k,n} \ba_{n-k+2} &= 0, \quad n \geq k-2,
     \label{eqrec0}
\end{align}
where
\begin{align}
    \ba_{n} \defi \begin{pmatrix}
        a^{(1)}_n \\
        a^{(2)}_n
    \end{pmatrix},
\end{align}
and $\bA_{1,n}, \dots, \bA_{k,n}$ are $2 \times 2$ matrices in terms of $\omega$ and black hole parameters.
The number of terms, $k$, depends on the potential in the master equation considered, but it can be reduced by the matrix-valued Gaussian elimination.
In fact, defining new matrices by
\begin{widetext}
\begin{align}
    \til{\bA}_{1,n} &\defi \bA_{1,n}, \quad n \geq 0,
    \label{eqmatA0}
    \\
    \til{\bA}_{2,n} &\defi
    \begin{cases}
        \bA_{2,n}, &0 \leq n \leq k-3,\\
        \bA_{2,n} - \bA_{k,n} \til{\bA}_{k-1,n-1}^{-1} \til{\bA}_{1,n-1}, &n \geq k-2,
    \end{cases} 
    \\
    \til{\bA}_{3,n} &\defi
    \begin{cases}
        \bA_{3,n}, &1 \leq n \leq k-3,\\
        \bA_{3,n} - \bA_{k,n} \til{\bA}_{k-1,n-1}^{-1} \til{\bA}_{2,n-1}, &n \geq k-2,
    \end{cases} 
    \\
    &\vdots\notag\\
    \til{\bA}_{k-1,n} &\defi
    \begin{cases}
        \bA_{k-1,n}, &n = k-3,\\
        \bA_{k-1,n} - \bA_{k,n} \til{\bA}_{k-1,n-1}^{-1} \til{\bA}_{k-2,n-1}, &n \geq k-2,
    \end{cases}
    \label{eqmatA}
\end{align}
\end{widetext}
one can show that the coefficient vectors $\ba_n$ satisfy a $(k-1)$-term recurrence relation in terms of the new matrices:
\begin{align}
    \til{\bA}_{1,0} \ba_{1} + \til{\bA}_{2,0} \ba_{0} &= 0,
    \\
    \til{\bA}_{1,1} \ba_{2} + \til{\bA}_{2,1} \ba_{1} + \til{\bA}_{3,1} \ba_{0} &= 0,
    \\
    &\vdots
    \notag \\
    \til{\bA}_{1,k-4} \ba_{k-3} + \til{\bA}_{2,k-4} \ba_{k-4} &
    \notag \\
    + \til{\bA}_{3,k-4} \ba_{k-5} + \cdots + \til{\bA}_{k-2,k-4} \ba_{0} &= 0, 
    \\
    \til{\bA}_{1,n} \ba_{n+1} + \til{\bA}_{2,n} \ba_{n} + \til{\bA}_{3,n} \ba_{n-1} &
    \notag \\
     + \cdots + \til{\bA}_{k-1,n} \ba_{n-k+3} &= 0, \quad n \geq k - 3.
\end{align}
Repeating the eliminations, the recurrence relation eventually settles into a three-term one,
\begin{align}
    \bA_{1,0} \ba_{1} + \bA_{2,0} \ba_{0} &= 0,
    \label{eqrec1}
    \\
    \bA_{1,n} \ba_{n+1} + \bA_{2,n} \ba_{n} + \bA_{3,n} \ba_{n-1} &= 0, \quad n \geq 1
    \label{eqrec2}
\end{align}
where we redefined finally resulting matrices via a repeat of Eqs.\,\eqref{eqmatA0}--\eqref{eqmatA} as $\bA_{1,n}$, $\bA_{2,n}$, and $\bA_{3,n}$.
Let us introduce a ladder matrix $\bR^+_n$ such that
\begin{align}
    \ba_{n+1} = \bR^+_n \ba_n.
    \label{eqlad}
\end{align}
Then, Eq.\,\eqref{eqrec2} leads to a recurrence relation for $\bR^+_n$,
\begin{align}
    \bR^+_n = - \left( \bA_{2,n+1} + \bA_{1,n+1} \bR^+_{n+1} \right)^{-1} \bA_{3,n+1},
    \label{eqladrec}
\end{align}
and Eq.\,\eqref{eqrec1} is reduced to
\begin{align}
    \bM \ba_0 = 0, \quad
    \bM \defi \bA_{2,0} + \bA_{1,0} \bR^+_0.
\end{align}
Finally, by requiring the existence of nontrivial eigenfunctions corresponding to quasinormal modes, the quasinormal frequencies can be found as $\omega$ satisfying
\begin{align}
    \det \bM = 0.
\end{align} 
We should note that $\bM$ can be expressed as the matrix-valued continued fractions by means of the recurrence relation \eqref{eqladrec}:
\begin{align}
    \bM &= \bA_{2,0} + \bA_{1,0} \bR^+_0
    \notag \\
    &= \bA_{2,0} - \bA_{1,0} ( \bA_{2,1} + \bA_{1,1} \bR^+_{1} )^{-1} \bA_{3,1}
    \notag \\
    &=  \bA_{2,0} - \bA_{1,0}
    \notag \\
    &\quad \times ( \bA_{2,1} - \bA_{1,1} ( \bA_{2,2} + \bA_{1,2} \bR^+_{2} )^{-1} \bA_{3,2} )^{-1} \bA_{3,1}
    \notag \\
    &= \cdots.
\end{align}
Practically, we can truncate the substitution of Eq.\,\eqref{eqladrec} at some large number of times $N$, and take $\bR^+_N$ to be arbitrary.
The $N$ must be large enough so that the obtained quasinormal frequencies converge.

The continued fractions method is useful when the potential in the master equation is given by powers of $1/r$.
To take advantage of the method, we expand the expressions of the potential components \eqref{eqEFTVI11}--\eqref{eqEFTVII22} with respect to $\bar{\alpha}$ and $\bar{\beta}$, and pick up only the terms up to the linear order of $\bar{\alpha}$ or $\bar{\beta}$.
Moreover, we use Eq.\,\eqref{eqrhexp} to approximate $r_\rh$ up to the linear order of $\bar{\alpha}$ for simplicity.
These approximations are valid for finding the leading corrections from nonlinear electrodynamics, which is our main purpose.

Under the above approximations, we find that the recurrence relations \eqref{eqrec0} are firstly given by a $k = 9$-term one for the type I, and a $k = 13$-term one for the type II.
The explicit expressions are terribly long, so we do not write them down here, but one can obtain them by direct insertions.
We take the order of continued fractions to be large enough so that the results converge within the desired precision, which is typically $N \sim 10^1$--$10^2$.
Given parameters $Q$, $\bar{\alpha}$, and $\bar{\beta}$ (taking $GM$ as the unit), we find quasinormal frequencies, i.e., zeros of $\det \bM$, by a numerical way such as the Newton's method.

\subsection{Results}
\label{subsec:results}

Here we show calculation results for the quasinormal frequencies by means of the methods in Sec.\,\ref{subsec:calculation}.
In particular, we present only fundamental frequencies in $l=2,3$ modes, which are expected to be observationally important.
As mentioned above, the perturbations are separated into the type I and II according to parity, and in each type there are two families of quasinormal modes $Z_1$ and $Z_2$, with corresponding quasinormal frequencies $\omega_1$ and $\omega_2$, respectively.
In practical calculations, we always work with quantities in the unit of $GM$, such as $r/(GM)$, $Q/(GM)$, and $\omega_{1,2} GM$.
The calculations are performed for black holes with various charges $Q$ as varying perturbation parameters $\bar{\alpha}$ and $\bar{\beta}$.

\begin{figure*}
    \centering
    \includegraphics[height=80mm]{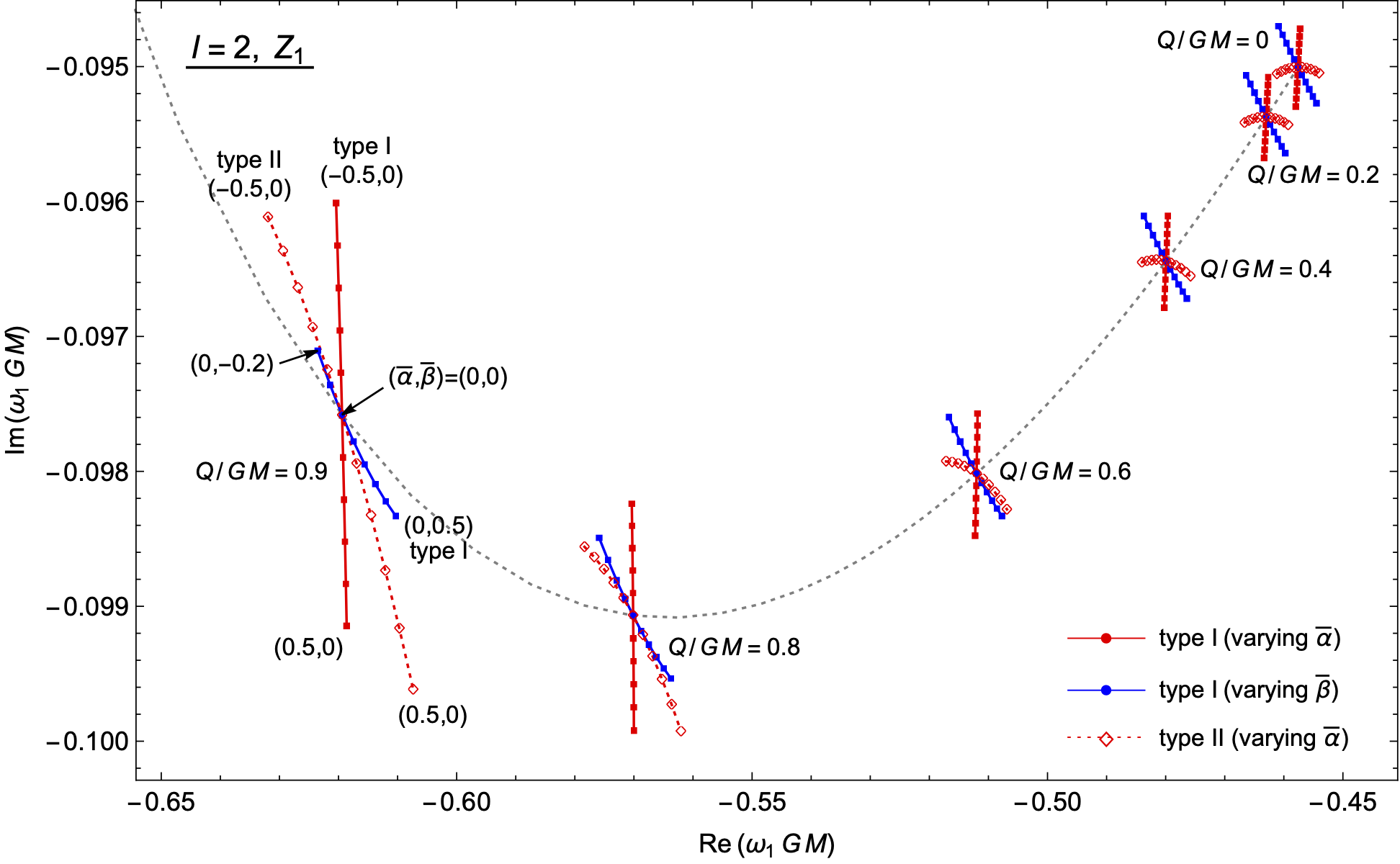}
    \caption{Fundamental quasinormal frequencies of the mode $Z_1$ of $l = 2$ for magnetic black holes with different charge $Q$, which are calculated by the numerical integration in Sec.\,\ref{subsubsec:numaric}. For different $Q$, the fundamental frequencies calculated as varying either $\bar{\alpha}$ or $\bar{\beta}$ are plotted: red solid lines denote the frequencies of the type I with $\bar{\alpha}$ varying from $-0.5$ to $0.5$ from top to bottom while $\bar{\beta}$ is fixed to be zero; blue solid lines denote the frequencies of the type I with $\bar{\beta}$ varying from $-0.5$ to $0.5$ for $Q/GM = 0, 0.2, 0.4, 0.6$, from $-0.4$ to $0.5$ for $Q/GM = 0.8$, and from $-0.2$ to $0.5$ for $Q/GM = 0.9$, from top-left to bottom-right while $\bar{\alpha}$ is fixed to be zero; red dashed lines denote the frequencies of the type II with $\bar{\alpha}$ varying from $-0.5$ to $0.5$ from left to right while $\bar{\beta}$ is fixed to be zero. The dots on the colored lines are marked for each 0.1 change in $\bar{\alpha}$ or $\bar{\beta}$. A gray dashed curve represents a series of quasinormal frequencies of Reissner--Nordstr\"om black holes ($\bar{\alpha} = \bar{\beta} = 0$).}
    \label{fig-l2Z1}
\end{figure*}

\begin{figure*}
    \centering
    \includegraphics[height=80mm]{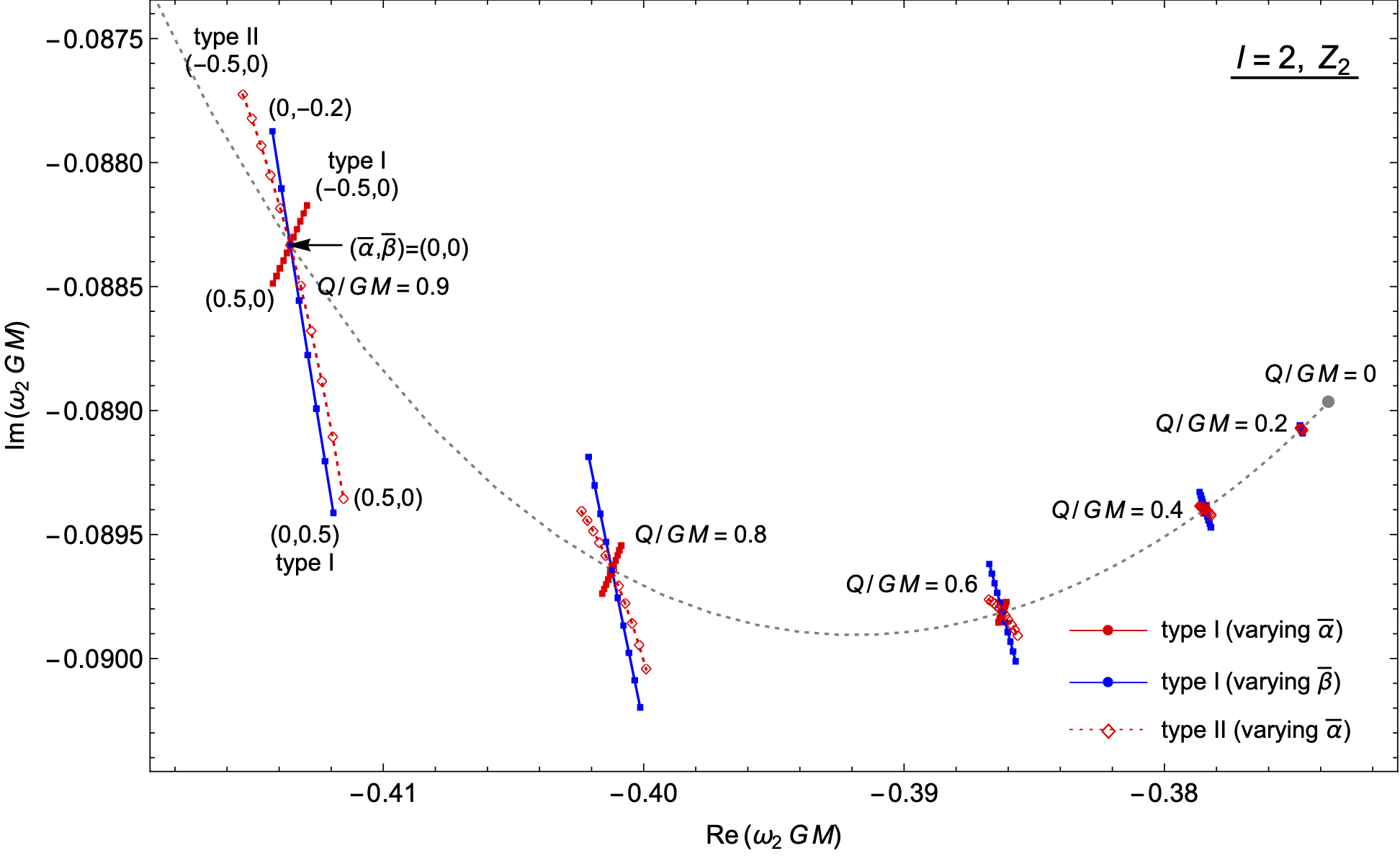}
    \caption{Fundamental quasinormal frequencies of the mode $Z_2$ of $l = 2$ for magnetic black holes with different charge $Q$. Descriptions of lines are the same as in Fig.\,\ref{fig-l2Z1}.}
    \label{fig-l2Z2}
\end{figure*}

\begin{figure*}
    \centering
    \includegraphics[height=80mm]{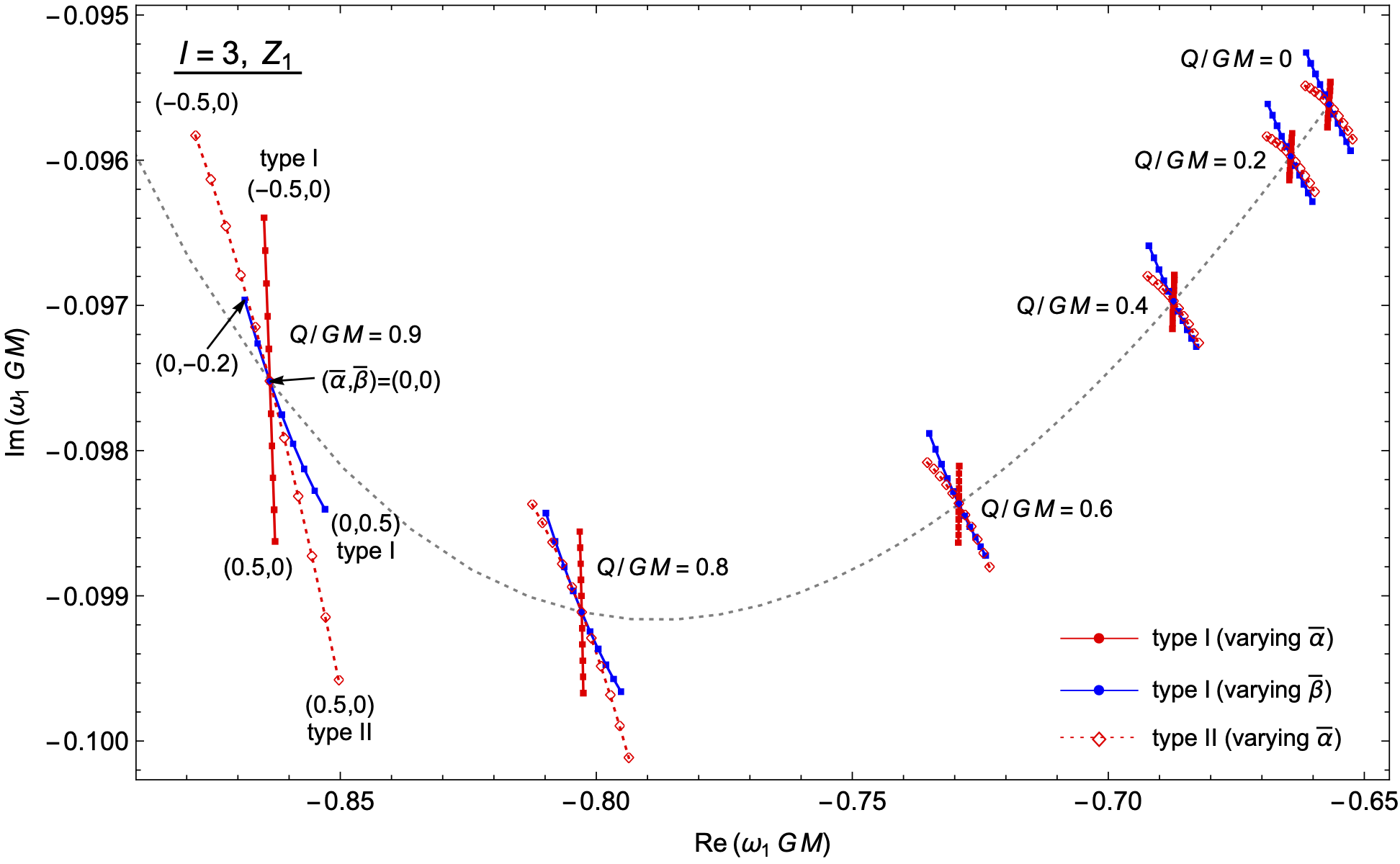}
    \caption{Fundamental quasinormal frequencies of the mode $Z_1$ of $l = 3$ for magnetic black holes with different charge $Q$. Descriptions of lines are the same as in Fig.\,\ref{fig-l2Z1}.}
    \label{fig-l3Z1}
\end{figure*}

\begin{figure*}
    \centering
    \includegraphics[height=80mm]{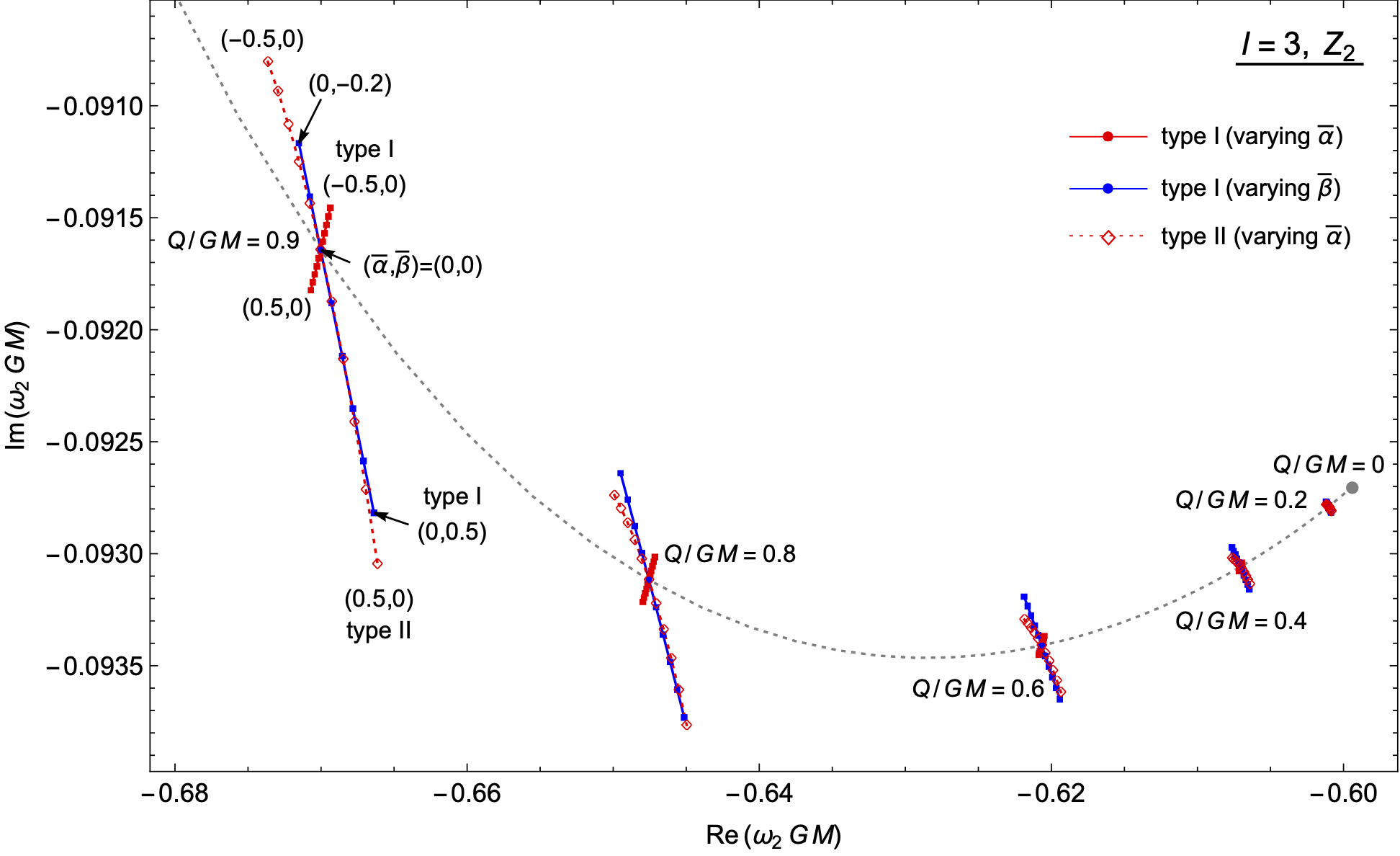}
    \caption{Fundamental quasinormal frequencies of the mode $Z_2$ of $l = 3$ for magnetic black holes with different charge $Q$. Descriptions of lines are the same as in Fig.\,\ref{fig-l2Z1}.}
    \label{fig-l3Z2}
\end{figure*}

The fundamental quasinormal frequencies calculated by the numerical integration in Sec.\,\ref{subsubsec:numaric} are plotted in figures: Fig.\,\ref{fig-l2Z1} is for the mode $Z_1$ of $l = 2$, Fig.\,\ref{fig-l2Z2} is for the mode $Z_2$ of $l = 2$, Fig.\,\ref{fig-l3Z1} is for the mode $Z_1$ of $l = 3$, and Fig.\,\ref{fig-l3Z2} is for the mode $Z_2$ of $l = 3$.
For different $Q$, the results calculated as varying either $\bar{\alpha}$ or $\bar{\beta}$ (while the other is fixed to be zero) are plotted as colored lines. 
Specifically, in each figure, red solid lines denote the frequencies of the type I with $\bar{\alpha}$ varying from $-0.5$ to $0.5$ while $\bar{\beta}$ is fixed to be zero, blue solid lines denote the frequencies of the type I with $\bar{\beta}$ varying from $-0.5$ to $0.5$ except for $Q/GM = 0.8$ and $0.9$ while $\bar{\alpha}$ is fixed to be zero, and red dashed lines denote the frequencies of the type II with $\bar{\alpha}$ varying from $-0.5$ to $0.5$ while $\bar{\beta}$ is fixed to be zero. 
(Only the ranges $-0.4 \leq \bar{\beta} \leq 0.5$ and $-0.2 \leq \bar{\beta} \leq 0.5$ are plotted for $Q/GM = 0.8$ and $0.9$, respectively, for the technical reason that our numerical integration cannot give results with desired accuracy when $\bar{\beta}$ is negatively too large.)
Note that the quasinormal frequencies in the type II are insensitive to $\bar{\beta}$ since the potential components \eqref{eqEFTVII11}--\eqref{eqEFTVII22} do not depend on $\bar{\beta}$.
Also note that ``$Q/GM=0$'' in the figures indicates the case with extremely small charge $Q$ compared to $GM$:
we calculate this case bearing in mind that, even if $Q/GM$ is extremely small, the parameters $\bar{\alpha}$ and $\bar{\beta}$ can take finite values to give nontrivial corrections depending on the black hole mass since our definitions of $\bar{\alpha}$ and $\bar{\beta}$, Eqs.\,\eqref{eqpatalpha} and \eqref{eqpatbeta}, are rewritten as
\begin{align}
    \bar{\alpha} &= \frac{\alpha}{4\pi G(GM)^2} \left(\frac{Q}{GM}\right)^2, 
    \notag \\
    \bar{\beta} &= \frac{\beta}{4\pi G(GM)^2} \left(\frac{Q}{GM}\right)^2\,.
    \label{eqpetpar}
\end{align}
When $Q/GM$ is infinitely small, the off-diagonal components of the potential matrix, $V_{\mathrm{I/II},12} = V_{\mathrm{I/II},21}$ in Eqs.\,\eqref{eqEFTVI12} and \eqref{eqEFTVII12}, vanish so that the master equations become decoupled. 
Furthermore, as $Q / GM \to 0$, the metric component $f(r)$ and potential component $V_{\mathrm{I}/\mathrm{II},11}$ lose the dependence on $\bar{\alpha}$ and $\bar{\beta}$, see Eqs.\,\eqref{eqEFTbg}, \eqref{eqEFTVI11} and \eqref{eqEFTVII11}.
Then, the master variable $\cR$ purely represents the gravitational perturbation on the Schwarzschild black hole, which corresponds to the mode $Z_2$ of ``$Q/GM=0$'' shown in Figs.\,\ref{fig-l2Z2} and \ref{fig-l3Z2}.
On the other hand, even if $Q / GM$ is extremely small, the $\bar{\alpha}$ and $\bar{\beta}$-dependence of the potentials $V_{\mathrm{I}/\mathrm{II},22}$ does not vanish explicitly, see Eqs.\,\eqref{eqEFTVI22} and \eqref{eqEFTVII22}. Therefore, we can obtain the nontrivial correction for the electromagnetic quasinormal mode $\cE$ due to $\bar{\alpha}$ and $\bar{\beta}$ even in the limit $Q/GM \to 0$, which corresponds to $Z_1$ of ``$Q/GM=0$'' shown in Figs.\,\ref{fig-l2Z1} and \ref{fig-l3Z1}.


From Figs.\,\ref{fig-l2Z1}--\ref{fig-l3Z2}, it turns out that the quasinormal frequencies can be linearly approximated with respect to $\bar{\alpha}$ and $\bar{\beta}$, at least for sufficiently small $\bar{\alpha}$ and $\bar{\beta}$. 
Let us parameterize the quasinormal frequencies as
\begin{align}
    \omega = \omega_{\text{RN}} + \bar{\alpha} \kappa_\alpha + \bar{\beta} \kappa_\beta
    \label{eqfit}
\end{align}
for small $\bar{\alpha}$ and $\bar{\beta}$, where $\kappa_\alpha$ and $\kappa_\beta$ are complex coefficients.
Then $\omega_{\text{RN}}$ corresponds to the case without nonlinear electrodynamics correction, i.e., one of the Reissner--Nordstr\"om black hole.
The coefficients $\kappa_\alpha$ and $\kappa_\beta$ can be obtained by fitting the calculated frequencies with respect to $\bar{\alpha}$ and $\bar{\beta}$, which are shown in Table \ref{tab-l2} for $l=2$, and in Table \ref{tab-l3} for $l=3$. We find that the results for the frequencies obtained by the numerical integration in Sec.\,\ref{subsubsec:numaric} and those by the continued fractions method in Sec.\,\ref{subsubsec:fractions} are in agreement within the desired precision.
Although we take the linear approximation for the master equations with respect to $\bar{\alpha}$ and $\bar{\beta}$ in the continued fractions method for simplicity as mentioned in Sec.\,\ref{subsubsec:fractions}, it turns out that the approximation is valid for finding  $\kappa_\alpha$ and $\kappa_\beta$ with sufficient accuracy.

From these results, we can easily read off the violation of isospectrality between type I and II due to nonlinear electrodynamics: quasinormal frequencies of the type I and II coincide for each charge $Q$ in the Reissner--Nordstr\"om ($\bar{\alpha} = \bar{\beta} = 0$) case, but this is no longer true once $\bar{\alpha}$ or $\bar{\beta}$ is non-vanishing.
Our analysis confirms previous results in Ref.\,\cite{Chaverra:2016ttw} where the violation of isospectrality has been found for black holes in a certain theory of nonlinear electrodynamics.

\begin{table*}
    \begin{tabular}{|l|l|l|l|l|l|l|l|l|l|l|} \hline
    $l=2$&\multicolumn{5}{c|}{$Z_1$} &\multicolumn{5}{c|}{$Z_2$} \\
    &       & \multicolumn{2}{c|}{type I} & \multicolumn{2}{c|}{type II} & & \multicolumn{2}{c|}{type I} & \multicolumn{2}{c|}{type II} \\
    $Q/GM$ &$\omega_{\text{RN}}GM$&$\kappa_\alpha GM$ &$\kappa_\beta GM$ &$\kappa_\alpha GM$ &$\kappa_\beta GM$ &$\omega_{\text{RN}} GM$ &$\kappa_\alpha GM$ &$\kappa_\beta GM$  &$\kappa_\alpha GM$   &$\kappa_\beta GM$ \\ \hline
    0   &-0.45760   &-0.00071   &0.00643    &0.00713    &0      &-0.37367   &0          &0    &0  &0  \\
        &-0.09500i  &-0.00058i  &-0.00057i  &+0.00001i  &+0i    &-0.08896i  &+0i        &+0i    &+0i    &+0i  \\\hline
    0.2 &-0.46297   &-0.00068   &0.00660    &0.00735    &0      &-0.37474   &-0.00003   &0.00011    &0.00012    &0   \\
        &-0.09537i  &-0.00060i  &-0.00058i  &-0.00001i  &+0i    &-0.08907i  &-0.00001i  &-0.00003i  &-0.00001i  &+0i \\\hline
    0.4 &-0.47993   &-0.00061   &0.00727    &0.00819    &0      &-0.37844   &-0.00012   &0.00043    &0.00046    &0   \\
        &-0.09644i  &-0.00068i  &-0.00061i  &-0.00010i  &+0i    &-0.08940i  &-0.00003i  &-0.00014i  &-0.00003i  &+0i \\\hline
    0.6 &-0.51201   &-0.00041   &0.00896    &0.01032    &0      &-0.38622   &-0.00030   &0.00101    &0.00111    &0   \\
        &-0.09802i  &-0.00091i  &-0.00073i  &-0.00035i  &+0i    &-0.08981i  &-0.00008i  &-0.00039i  &-0.00014i  &+0i \\\hline
    0.8 &-0.57013   &0.00036    &0.01363    &0.01637    &0      &-0.40122   &-0.00072   &0.00220    &0.00249    &0   \\
        &-0.09907i  &-0.00168i  &-0.00119i  &-0.00135i  &+0i    &-0.08964i  &-0.00019i  &-0.00113i  &-0.00061i  &+0i \\\hline
    0.9 &-0.61940   &0.00180    &0.01980    &0.02469    &0      &-0.41357   &-0.00130   &0.00340    &0.00391    &0 \\
        &-0.09758i  &-0.00314i  &-0.00208i  &-0.00349i  &+0i    &-0.08833i   &-0.00032i  &-0.00226i  &-0.00156i  &+0i \\\hline
    \end{tabular}
    \caption{$\omega_{\mathrm{RN}}$, $\kappa_\alpha$ and $\kappa_\beta$ in Eq.\,\eqref{eqfit} for fundamental quasinormal frequencies of the mode $l=2$ of black holes with different charge $Q$.}
    \label{tab-l2}
\end{table*}

\begin{table*}
    \begin{tabular}{|l|l|l|l|l|l|l|l|l|l|l|} \hline
    $l=3$&\multicolumn{5}{c|}{$Z_1$} &\multicolumn{5}{c|}{$Z_2$} \\
    &       & \multicolumn{2}{c|}{type I} & \multicolumn{2}{c|}{type II} & & \multicolumn{2}{c|}{type I} & \multicolumn{2}{c|}{type II} \\
    $Q/GM$ &$\omega_{\text{RN}}GM$&$\kappa_\alpha GM$ &$\kappa_\beta GM$ &$\kappa_\alpha GM$ &$\kappa_\beta GM$ &$\omega_{\text{RN}} GM$ &$\kappa_\alpha GM$ &$\kappa_\beta GM$  &$\kappa_\alpha GM$   &$\kappa_\beta GM$ \\ \hline
    0   &-0.65690   &-0.00059   &0.00868    &0.00927    &0      &-0.59944   &0          &0      &0      &0  \\
        &-0.09562i  &-0.00031i &-0.00067i   &-0.00036i  &+0i    &-0.09270i  &+0i    &+0i        &+0i    &+0i  \\\hline
    0.2 &-0.66437   &-0.00054   &0.00872    &0.00935    &0      &-0.60103   &-0.00005   &0.00035    &0.00035    &0   \\
        &-0.09597i  &-0.00032i  &-0.00067i  &-0.00038i  &+0i    &-0.09279i  &-0.00001i  &-0.00005i  &-0.00003i  &+0i \\\hline
    0.4 &-0.68728   &-0.00041   &0.00921    &0.00998    &0      &-0.60706   &-0.00018   &0.00118    &0.00121    &0   \\
        &-0.09697i  &-0.00037i  &-0.00069i  &-0.00046i  &+0i    &-0.09306i  &-0.00004i  &-0.00019i  &-0.00011i  &+0i \\\hline
    0.6 &-0.72919   &-0.00015   &0.01101    &0.01218    &0      &-0.62066   &-0.00039   &0.00244    &0.00250    &0   \\
        &-0.09837i  &-0.00053i  &-0.00083i  &-0.00073i  &+0i    &-0.09341i  &-0.00008i  &-0.00046i  &-0.00031i  &+0i \\\hline
    0.8 &-0.80284   &0.00069    &0.01644    &0.01888    &0      &-0.64755   &-0.00082   &0.00484    &0.00497    &0   \\
        &-0.09911i  &-0.00111i  &-0.00140i  &-0.00176i  &+0i    &-0.09312i  &-0.00020i  &-0.00121i  &-0.00100i  &+0i \\\hline
    0.9 &-0.86376   &0.00216    &0.02365    &0.02807    &0      &-0.67002   &-0.00133   &0.00741    &0.00758    &0 \\
        &-0.09752i  &-0.00223i  &-0.00245i  &-0.00381i  &+0i    &-0.09164i  &-0.00037i  &-0.00238i  &-0.00219i  &+0i \\\hline
    \end{tabular}
    \caption{$\omega_{\mathrm{RN}}$, $\kappa_\alpha$ and $\kappa_\beta$ in Eq.\,\eqref{eqfit} for fundamental quasinormal frequencies of the mode $l=3$ of black holes with different charge $Q$.}
    \label{tab-l3}
\end{table*}

\subsection{Applications}
\label{subsec:Applications}

So far, we have treated $\alpha$ and $\beta$ in the effective Lagrangian \eqref{eqeff} as free parameters, and derived quasinormal frequencies of black holes with various charges in terms of $\alpha$ and $\beta$. By specifying $\alpha$ and $\beta$, we can obtain quasinormal frequencies in various theories of nonlinear electrodynamics. Below, as examples of the application, we calculate the quasinormal frequencies in two nonlinear electrodynamics: the Euler--Heisenberg theory, and the Born--Infeld theory.

\subsubsection{Euler--Heisenberg electrodynamics}
\label{subsubsec:eulerheisenberg}

First, let us find the corrections for quasinormal frequencies due to the Euler--Heisenberg nonlinear electrodynamics taking in electron one-loop corrections \cite{Heisenberg:1936nmg}.
Up to the quadratic order of $\cF$ and $\cG$, the Euler--Heisenberg Lagrangian is given by
\begin{align}
    \cL(\cF, \cG) = \cF - \frac{2}{45 m_e^4} \left( \frac{e^2}{4 \pi} \right)^2 (4 \cF^2 + 7 \cG^2)\,,
\end{align}
where $m_e$ is the mass of the electron and $e^2/(4\pi)$ is the fine structure constant.
In this case, the perturbation parameters introduced in Eqs.\,\eqref{eqpatalpha} and \eqref{eqpatbeta} read
\begin{align}
    \bar{\alpha} &= \frac{8}{45 m_e^4} \left(\frac{e^2}{4\pi}\right)^2 \frac{1}{4 \pi G^3 M^2} \left(\frac{Q}{GM} \right)^2 , \quad
    \bar{\beta} = \frac{7}{4} \bar{\alpha}.
    \label{eqEHpat}
\end{align}
First of all, the parameter $\bar{\alpha}$ should be sufficiently small to ensure the validity of the effective theory.
Inserting the physical constants, $\bar{\alpha}$ in Eq.\,\eqref{eqEHpat} can be rewritten as
\begin{align}
    \bar{\alpha} = 2.9 \times 10^{-5} \left(\frac{10^6 M_\odot}{M}\right)^2 \left(\frac{Q}{GM}\right)^2.
    \label{eqalphaEH}
\end{align}
Therefore, in the case of a charged supermassive black hole, we can expect that the Euler--Heisenberg electrodynamics effectively holds around the horizon, and we can calculate the corrections for the quasinormal modes.
Let us linearly approximate the fundamental quasinormal frequencies with the Euler--Heisenberg corrections as
\begin{align}
    \omega = \omega_{\text{RN}} + \bar{\alpha} \kappa_{\text{EH}}\,,
    \label{eqfreEH}
\end{align}
for small $\bar{\alpha}$. Here, the coefficient $\kappa_{\text{EH}}$ can be found as 
\begin{align}
    \kappa_{\text{EH}} = \kappa_{\alpha} + \frac{7}{4} \kappa_{\beta}
\end{align}
in terms of $\kappa_{\alpha}$ and $\kappa_{\beta}$ given by Table \ref{tab-l2} for $l=2$, and Table \ref{tab-l3} for $l=3$.
We explicitly show the values of $\kappa_{\text{EH}}$ in Table \ref{tab-EHl2} for $l=2$, and in Table \ref{tab-EHl3} for $l=3$.
From the tables, it can be seen that the Euler--Heisenberg corrections reduce the oscillation frequency $|\text{Re}(\omega)|$ and enhance the damping rate $|\text{Im}(\omega)|$ compared to the linear Maxwell theory, except for the mode $Z_1$ in the type II of $l=2$ for $Q/GM = 0$.
In other words, by the Euler--Heisenberg corrections, the oscillation period of the quasinormal modes become longer, and the black hole settles into a stationary state more quickly, in almost all cases.
The decreasing of $|\text{Re}(\omega)|$ can be understood from the lowering of the height of the effective potential in the master equations as drawn in Fig.\,\ref{fig-pot}.
This tendency of the oscillation frequency and damping rate due to the Euler--Heisenberg corrections is consistent with the result of Ref.\,\cite{Breton:2021mju}, where the quasinormal modes of Euler--Heisenberg black holes are studied by the eikonal approximation.

\begin{table}[t]
    \begin{tabular}{|l|l|l|l|l|} \hline
    EH, $l=2$&\multicolumn{2}{c|}{$Z_1$} &\multicolumn{2}{c|}{$Z_2$} \\
    &type I &type II &type I&type II\\
    $Q/GM$ &$\kappa_{\text{EH}} GM$ &$\kappa_{\text{EH}} GM$ &$\kappa_{\text{EH}} GM$ &$\kappa_{\text{EH}} GM$ \\ \hline
    0   &0.01054	&0.00713	&0      	&0      \\
        &-0.00157i	&+0.00001i	&+0i	    &+0i    \\\hline
    0.2 &0.01087	&0.00735	&0.00016	&0.00012\\
        &-0.00161i	&-0.00001i	&-0.00007i	&-0.00001i\\\hline
    0.4 &0.01211	&0.00819	&0.00062	&0.00046\\
        &-0.00175i	&-0.00010i	&-0.00028i	&-0.00003i\\\hline
    0.6 &0.01528	&0.01032	&0.00147	&0.00111\\
        &-0.00218i	&-0.00035i	&-0.00077i	&-0.00014i\\\hline
    0.8 &0.02422	&0.01637	&0.00312	&0.00249\\
        &-0.00376i  &-0.00135i	&-0.00217i	&-0.00061i\\\hline
    0.9 &0.03644	&0.02469	&0.00466	&0.00391\\
        &-0.00678i	&-0.00349i	&-0.00427i	&-0.00156i\\\hline
    \end{tabular}
    \caption{The correction coefficients due to Euler--Heisenberg electrodynamics $\kappa_{\text{EH}}$ in Eq.\,\eqref{eqfreEH} for fundamental quasinormal frequencies of the mode $l=2$ of black holes with different charge $Q$.}
    \label{tab-EHl2}
\end{table}

\begin{table}[t]
    \begin{tabular}{|l|l|l|l|l|} \hline
    EH, $l=3$&\multicolumn{2}{c|}{$Z_1$} &\multicolumn{2}{c|}{$Z_2$} \\
    &type I &type II &type I&type II\\
    $Q/GM$ &$\kappa_{\text{EH}} GM$ &$\kappa_{\text{EH}} GM$ &$\kappa_{\text{EH}} GM$ &$\kappa_{\text{EH}} GM$ \\ \hline
    0   &0.01461	&0.00927	&0	    &0\\
        &-0.00149i	&-0.00036i	&+0i	&+0i\\\hline
    0.2 &0.01472	&0.00935	&0.00055	&0.00035\\
        &-0.00149i	&-0.00038i	&-0.00010i	&-0.00003i\\\hline
    0.4 &0.01571	&0.00998	&0.00189	&0.00121\\
        &-0.00158i	&-0.00046i	&-0.00036i	&-0.00011i\\\hline
    0.6 &0.01912	&0.01218	&0.00388	&0.00250\\
        &-0.00199i	&-0.00073i	&-0.00089i	&-0.00031i\\\hline
    0.8 &0.02947	&0.01888	&0.00766	&0.00497\\
        &-0.00357i	&-0.00176i	&-0.00232i	&-0.00100i\\\hline
    0.9 &0.04355	&0.02807	&0.01164	&0.00758\\
        &-0.00652i	&-0.00381i	&-0.00453i	&-0.00219i\\\hline
    \end{tabular}
    \caption{The correction coefficients due to Euler--Heisenberg electrodynamics $\kappa_{\text{EH}}$ in Eq.\,\eqref{eqfreEH} for fundamental quasinormal frequencies of the mode $l=3$ of black holes with different charge $Q$.}
    \label{tab-EHl3}
\end{table}

For example, if we consider the quasinormal mode of $l=2$ on a charged black hole with $Q/(GM) =0.9$, the Euler--Heisenberg corrections contribute as
\begin{align}
    Z_1: &\frac{\text{Re}(\bar{\alpha} \kappa_{\text{EH}})}{\text{Re}(\omega_{\text{RN}})}
    =
    \begin{cases} -1.4 \times 10^{-6} \left({10^6 M_\odot}/{M} \right)^2
    & (\text{type I}),
     \\
    -9.5 \times 10^{-7} \left({10^6 M_\odot}/{M} \right)^2
    & (\text{type II}),
    \end{cases}
    \notag\\
    &\frac{\text{Im}(\bar{\alpha} \kappa_{\text{EH}})}{\text{Im}(\omega_{\text{RN}})}
    =
    \begin{cases} 1.7 \times 10^{-6} \left({10^6 M_\odot}/{M} \right)^2
    & (\text{type I}),
     \\
    8.5 \times 10^{-7} \left({10^6 M_\odot}/{M} \right)^2
    & (\text{type II}),
    \end{cases}
    \notag\\
    Z_2: &\frac{\text{Re}(\bar{\alpha} \kappa_{\text{EH}})}{\text{Re}(\omega_{\text{RN}})}
    =
    \begin{cases} -2.7 \times 10^{-7} \left({10^6 M_\odot}/{M} \right)^2
    & (\text{type I}),
     \\
    -2.3 \times 10^{-7} \left({10^6 M_\odot}/{M} \right)^2
    & (\text{type II}),
    \end{cases}
    \notag\\
    &\frac{\text{Im}(\bar{\alpha} \kappa_{\text{EH}})}{\text{Im}(\omega_{\text{RN}})}
    =
    \begin{cases} 1.2 \times 10^{-6} \left({10^6 M_\odot}/{M} \right)^2
    & (\text{type I}),
     \\
    4.2 \times 10^{-7} \left({10^6 M_\odot}/{M} \right)^2
    & (\text{type II}),
    \end{cases}
    \notag\\
    &\hspace{7em}\text{for } l=2,~ \text{and } Q/GM = 0.9.
\end{align}
This implies that the Euler--Heisenberg electrodynamics taking in the electron one-loop corrections would yield $\sim 0.1$--$1\%$ modifications for the quasinormal frequencies of a black hole with $M \sim 10^4 M_\odot$ and $Q/(GM) = 0.9$, if it exists.

There are some discussions on a possible electric charge of realistic black holes. In Ref.\,\cite{Zajacek:2019kla}, assuming that the number of electrons and of protons are balanced under the gravitational and electric potentials, the plausible electric charge of the black hole is estimated. The resulting charge-to-mass ratio of the black hole can be written as 
\begin{align}
    \frac{Q}{GM} &= 2 \pi \left( \frac{e^2}{4\pi} \right)^{-1/2} \sqrt{G} (m_p - m_e)
    \notag \\
    &= 5.7 \times 10^{-18},
\end{align}
where $m_p$ is the mass of the proton. This ratio is extremely small, thus the possible corrections to the quasinormal frequencies can be estimated by utilizing the row of $Q / GM = 0$ in the tables. However, in this case, $\bar{\alpha}$ given by Eq.\,\eqref{eqalphaEH} would also be extremely small for astrophysical black holes, thus we can expect no significant corrections from the observational point of view.

\subsubsection{Born--Infeld electrodynamics}
\label{subsubsec:borninfeld}

As another example, let us apply our results to Born--Infeld nonlinear electrodynamics, which was originally proposed to remove the divergence of the self-energy of a charged particle within classical electrodynamics \cite{Born:1934gh}.
The complete Lagrangian is given by
\begin{align}
    \cL = \mu^4 \sqrt{ 1 + \frac{2\cF}{\mu^4} - \frac{\cG^2}{\mu^8}} - \mu^4,
\label{eqBI}
\end{align}
where $\mu$ is a parameter with a mass dimension. Then, the upper bound for the electric field strength is set as $\mu^2$ so that the self-energy of a charged particle becomes finite.
By expanding the Lagrangian \eqref{eqBI} up to the quadratic order in $\cF$ and $\cG$, we have 
\begin{align}
    \cL = \cF - \frac{\cF^2}{2\mu^4} - \frac{\cG^2}{2\mu^4}\,.
    \label{eqBI2}
\end{align}
This case corresponds to 
\begin{align}
    \alpha = \beta = \frac{1}{2\mu^4}\,
\end{align}
in the effective Lagrangian \eqref{eqeff}.
To ensure the validity of the perturbative expansion, we expect
\begin{align}
    \bar{\alpha} = \frac{1}{8\pi G \mu^4 (GM)^2} \left(\frac{Q}{GM}\right)^2 \ll 1\,.
    \label{eqalphaBI}
\end{align}
For small $\bar{\alpha}$, we can linearly approximate the quasinormal frequencies with the Born--Infeld corrections as
\begin{align}
    \omega = \omega_{\mathrm{RN}} + \bar{\alpha} \kappa_{\mathrm{BI}}\,,
    \label{eqfreBI}
\end{align}
where $\kappa_{\mathrm{BI}}$ is given by 
\begin{align}
    \kappa_{\mathrm{BI}} = \kappa_{\alpha} + \kappa_{\beta}\,,
    \label{eqcoeffBI}
\end{align}
in terms of $\kappa_\alpha$ and $\kappa_\beta$, which are obtained in the previous section for fundamental modes.
The values of $\kappa_{\mathrm{BI}}$ derived from Tables \ref{tab-l2} and \ref{tab-l3} are shown in Table \ref{tab-BIl2} for $l=2$, and in Table \ref{tab-BIl3} for $l=3$.
From the tables, we can see that the Born--Infeld electrodynamics contributes to reduce the oscillation frequency $|\text{Re}(\omega)|$ and enhance the damping rate $|\text{Im}(\omega)|$ compared to the linear Maxwell electrodynamics in almost all cases, as in the Euler--Heisenberg case.

\begin{table}[t]
    \begin{tabular}{|l|l|l|l|l|} \hline
    BI, $l=2$&\multicolumn{2}{c|}{$Z_1$} &\multicolumn{2}{c|}{$Z_2$} \\
    &type I &type II &type I&type II\\
    $Q/GM$ &$\kappa_{\text{BI}} GM$ &$\kappa_{\text{BI}} GM$ &$\kappa_{\text{BI}} GM$ &$\kappa_{\text{BI}} GM$ \\ \hline
    0   &0.00572	&0.00713	&0          &0      \\
        &-0.00115i	&+0.00001i	&+0i        &+0i    \\\hline
    0.2 &0.00592	&0.00735	&0.00008    &0.00012 \\
        &-0.00117i	&-0.00001i	&-0.00004i	&-0.00001i\\\hline        
    0.4 &0.00666    &0.00819    &0.00030    &0.00046 \\
        &-0.00129i  &-0.00010i	&-0.00018i	&-0.00003i\\\hline        
    0.6 &0.00856    &0.01032	&0.00071    &0.00111 \\
        &-0.00163i  &-0.00035i	&-0.00048i	&-0.00014i\\\hline        
    0.8 &0.01399    &0.01637	&0.00148    &0.00249 \\
        &-0.00287i  &-0.00135i	&-0.00132i	&-0.00061i\\\hline        
    0.9 &0.02159    &0.02469	&0.00210   	&0.00391 \\
        &-0.00522i  &-0.00349i	&-0.00257i	&-0.00156i \\\hline        
    \end{tabular}
    \caption{The correction coefficients due to Born--Infeld electrodynamics $\kappa_{\text{BI}}$ in Eq.\,\eqref{eqfreBI} for fundamental quasinormal frequencies of the mode $l=2$ of black holes with different charge $Q$.}
    \label{tab-BIl2}
\end{table}

\begin{table}[t]
    \begin{tabular}{|l|l|l|l|l|} \hline
    BI, $l=3$&\multicolumn{2}{c|}{$Z_1$} &\multicolumn{2}{c|}{$Z_2$} \\
    &type I &type II &type I&type II\\
    $Q/GM$ &$\kappa_{\text{BI}} GM$ &$\kappa_{\text{BI}} GM$ &$\kappa_{\text{BI}} GM$ &$\kappa_{\text{BI}} GM$ \\ \hline
    0   &0.00809	&0.00927	&0	    &0\\
        &-0.00099i	&-0.00036i	&+0i	&+0i\\\hline
    0.2 &0.00818	&0.00935	&0.00029	&0.00035\\
        &-0.00099i	&-0.00038i	&-0.00006i	&-0.00003i\\\hline
    0.4 &0.00880	&0.00998	&0.00100	&0.00121\\
        &-0.00106i	&-0.00046i	&-0.00022i	&-0.00011i\\\hline
    0.6 &0.01086	&0.01218	&0.00205	&0.00250\\
        &-0.00136i	&-0.00073i	&-0.00054i	&-0.00031i\\\hline
    0.8 &0.01714	&0.01888	&0.00402	&0.00497\\
        &-0.00251i	&-0.00176i	&-0.00141i	&-0.00100i\\\hline
    0.9 &0.02581	&0.02807	&0.00608	&0.00758\\
        &-0.00468i	&-0.00381i	&-0.00275i	&-0.00219i\\\hline
    \end{tabular}
    \caption{The correction coefficients due to Born--Infeld electrodynamics $\kappa_{\text{BI}}$ in Eq.\,\eqref{eqfreBI} for fundamental quasinormal frequencies of the mode $l=3$ of black holes with different charge $Q$.}
    \label{tab-BIl3}
\end{table}

If we take $\mu$ and $M$ in Eq.\,\eqref{eqBI} to be the Planck scale $\Mpl = G^{-1/2} = 1.2 \times 10^{19} \, \text{GeV} = 2.2 \times 10^{-5} \, \text{g}$ as a reference, the parameter $\bar{\alpha}$ given by Eq.\,\eqref{eqalphaBI} can be written as
\begin{align}
    \bar{\alpha} = 0.040 \times \left(\frac{\Mpl}{M}\right)^2 \left(\frac{\Mpl}{\mu}\right)^4 \left(\frac{Q}{GM}\right)^2.
\end{align}
In particular, for the mode $l=2$ on a charged black hole with $Q/(GM) =0.9$ and $M = \Mpl$, the Born--Infeld corrections with $\mu = \Mpl$ contribute as
\begin{align}
    Z_1: &\frac{\text{Re}(\bar{\alpha} \kappa_{\text{BI}})}{\text{Re}(\omega_{\text{RN}})}
    =
    \begin{cases} -1.1 \times 10^{-3} 
    & (\text{type I}),
     \\
    -1.3 \times 10^{-3} 
    & (\text{type II}),
    \end{cases}
    \notag\\
    &\frac{\text{Im}(\bar{\alpha} \kappa_{\text{BI}})}{\text{Im}(\omega_{\text{RN}})}
    =
    \begin{cases} 1.7 \times 10^{-3} 
    & (\text{type I}),
     \\
    1.2 \times 10^{-3} 
    & (\text{type II}),
    \end{cases}
    \notag\\
    Z_2: &\frac{\text{Re}(\bar{\alpha} \kappa_{\text{BI}})}{\text{Re}(\omega_{\text{RN}})}
    =
    \begin{cases} -1.6 \times 10^{-4} 
    & (\text{type I}),
     \\
    -3.0 \times 10^{-4} 
    & (\text{type II}),
    \end{cases}
    \notag\\
    &\frac{\text{Im}(\bar{\alpha} \kappa_{\text{BI}})}{\text{Im}(\omega_{\text{RN}})}
    =
    \begin{cases} 9.4 \times 10^{-4}
    & (\text{type I}),
     \\
    5.7 \times 10^{-4} 
    & (\text{type II}),
    \end{cases}
    \notag\\
    &\text{for } l=2, M = \Mpl, Q/GM = 0.9, \mu = \Mpl.
\end{align}
In this case, the relative corrections to the quasinormal frequencies can reach $\sim 0.1\%$.

\section{Summary and discussion}
\label{sec:Summary}

In this paper, we analyzed quasinormal modes related to gravitational and electromagnetic perturbations of static and spherically symmetric charged black holes in general relativity with nonlinear electrodynamics. In particular, we considered the effective field theory in nonlinear electrodynamics given by Eq.\,\eqref{eqeff} as giving the leading corrections to the Maxwell electrodynamics. The coefficients $\alpha$ and $\beta$ in the effective Lagrangian \eqref{eqeff} were treated as arbitrary parameters throughout the analysis so that the framework of nonlinear electrodynamics was inclusively treated. As commented in Sec.\,\ref{sec:EFT} and Appendix \ref{app:FPdualityEFT}, note that a magnetically charged black hole and an electrically charged black hole can be treated in parallel up to the linear order of the parameters $\alpha$ and $\beta$. To obtain the quasinormal frequencies, we independently used two calculation methods, the numerical integration and the continued fractions method, which are useful even for a multi-component system as in our case. In particular, we calculated quasinormal frequencies of the fundamental modes for $l = 2, 3$, and linearly parametrized the corrections for the frequencies in terms of the perturbation parameters as Eq.\,\eqref{eqfit}. The results by the two calculation methods agree, and they are listed in Tables \ref{tab-l2} and \ref{tab-l3}. Given a certain nonlinear electrodynamics with specific $\alpha$ and $\beta$, and given the mass and charge of the black hole, the perturbation parameters are given by Eq.\,\eqref{eqpetpar}, and then the corrections for the fundamental quasinormal frequencies can be calculated immediately from Eq.\,\eqref{eqfit} and Tables \ref{tab-l2} and \ref{tab-l3}. As examples, we calculated the corrections in Euler--Heisenberg and Born--Infeld electrodynamics in Sec.\,\ref{subsec:Applications}. 

From Tables \ref{tab-l2} and \ref{tab-l3}, we can see that the imaginary part of the correction coefficients $\kappa_{\alpha}$ and $\kappa_{\beta}$ is negative in almost all cases. This means that if $\alpha$ and $\beta$ in the effective Lagrangian are positive, the magnitude of the damping rate of the quasinormal modes will be enhanced due to the nonlinear electrodynamics corrections. One should note that $\alpha$ and $\beta$ must be positive if the effective theory has an ultraviolet completion which satisfies the basic requirements of physics, such as unitarity, causality, and locality \cite{Adams:2006sv}.

The system of electromagnetic and gravitational perturbations of black holes is separated into two types according to parity. For the Reissner--Nordstr\"om black hole in Einstein--Maxwell theory, the isospectrality of quasinormal modes under parity holds, i.e., the quasinormal frequencies of the two types exactly coincide \cite{chandrasekhar1998mathematical}. On the other hand, the results of the present paper show that the isospectrality is violated due to nonlinear electrodynamics corrections. This isospectrality violation was found in a specific model of nonlinear electrodynamics giving a regular black hole in Ref.\,\cite{Chaverra:2016ttw}, but in this paper we have confirmed that the violation occurs in the general case. As pointed out in that paper, this isospectrality violation would be partly explained in the eikonal limit, in which the quasinormal modes in the large $l$ are associated with unstable circular null orbits. In nonlinear electrodynamics, light rays follow null geodesics in effective metrics which are different from the spacetime metric and depend on the background electromagnetic field. The fact that two different effective metrics are possible according to the polarization would be related to the violation of isospectrality under parity. In fact, two different quasinormal frequencies of the Euler--Heisenberg black hole in the eikonal limit have been studied in Ref.\,\cite{Breton:2021mju} based on the effective metrics, and the tendency of the corrections, such as the enhancement of the damping rate, is consistent with the results of the present paper.

Throughout this paper, we have concentrated on the case where the effective electrodynamics characterized by $(F_{\mu\nu} F^{\mu\nu})^2$ and $(F_{\mu\nu} \til{F}^{\mu\nu})^2$ gives the leading corrections for the quasinormal modes. More generally, it will be important to study quasinormal modes of charged black holes in effective theories with higher-order corrections including spacetime curvature, such as $R_{\mu\nu\rho\sigma}F^{\mu\nu}F^{\rho\sigma}$. Furthermore, there are Lagrangians of nonlinear electrodynamics without a power series expansion in the scalars $\cF$ and $\cG$ unlike the framework treated in the present paper. In fact, in Ref.\,\cite{Bandos:2020jsw}, a theory of nonlinear electrodynamics with electromagnetic duality invariance and conformal invariance has been found as an extension of Maxwell electrodynamics, which does not have such an expansion. Black holes or other gravitating objects in this theory have been studied in Refs.\,\cite{Flores-Alfonso:2020euz, BallonBordo:2020jtw, Flores-Alfonso:2020nnd, Bokulic:2021dtz, Zhang:2021qga}. It will be interesting to analyze the quasinormal modes of black holes in such a theory. We leave these issues as future works.

\begin{acknowledgments}

The authors thank Tomonori Inoue, Sota Sato, Jiro Soda, and Hirotaka Yoshino for fruitful discussions. K.N. was supported by JSPS KAKENHI Grant Number JP21J20600. D.Y. was supported by JSPS KAKENHI Grant Numbers JP19J00294 and JP20K14469.
    
\end{acknowledgments}

\appendix

\section{Magnetic/electric black holes and $FP$ duality}
\label{app:duality}

In this Appendix, we review the constructions of spherically symmetric magnetic or electric black hole solutions in nonlinear electrodynamics, and the relationship between them known as $FP$ duality.

We start with the action \eqref{eqaction}.
Taking variation of the action with respect to the metric yields the Einstein equations,
\begin{align}
    {G_{\mu}}^{\nu} = 8 \pi G{T_{\mu}}^{\nu}\,,
\end{align}
where the left-hand side is the Einstein tensor calculated from $g_{\mu\nu}$, while the right-hand side is the energy-momentum tensor of the electromagnetic field in nonlinear electrodynamics, 
\begin{align}
    {T_{\mu}}^{\nu} &= \cL_\cF F_{\mu\lambda} F^{\nu\lambda} + \delta_\mu^\nu (\cL_\cG \cG - \cL)
    \notag \\
    &= \cL_\cF F_{\mu\lambda} F^{\nu\lambda} + \cL_\cG F_{\mu\lambda} \til{F}^{\nu\lambda} - \delta_\mu^\nu \cL\,,
\end{align}
where $\cL_\cF = \pd \cL/ \pd \cF$ and $\cL_\cG = \pd \cL/ \pd \cG$.
On the other hand, by taking variation of the action with respect to $A_\mu$, we have the equations of motion for the electromagnetic field,
\begin{align}
    \nabla_{\mu} \left(\cL_\cF F^{\mu\nu} + \cL_\cG \til{F}^{\mu\nu} \right) = 0\,.
\end{align}
The electromagnetic field strength tensor also satisfies the Bianchi identity,
\begin{align}
    \nabla_{\mu} \til{F}^{\mu\nu} = 0\,.
\end{align}

To find static and spherically symmetric solutions, we put an ansatz for the metric as
\begin{align}
    g_{\mu\nu} dx^\mu dx^\nu = - f(r) dt^2 + \frac{1}{f(r)} dr^2 + r^2(d\theta^2 + \sin^2 \theta d\phi^2)\,.
    \label{eqssa}
\end{align}
Then, the components of the Einstein tensor read
\begin{align}
    {G_t}^t = {G_r}^r &= - \frac{1}{r^2} + \frac{f}{r^2}+ \frac{f'}{r} \,,\\
    {G_\theta}^\theta = {G_\phi}^\phi &= \frac{f''}{2} + \frac{f'}{r} \,,
\end{align}
where a prime denotes the derivative with respect to $r$.

\subsection{Magnetic black holes}
\label{app:magnetic}

We first consider static and spherically symmetric solutions with magnetic charge.
We take the magnetic field configuration as
\begin{align}
    \frac{1}{2}F_{\mu\nu} dx^\mu \wedge dx^\nu = q \sin \theta \,d\theta \wedge d\phi\,,
\end{align}
where $q$ is a constant corresponding to the magnetic charge.
On this magnetic configuration, we have
\begin{align}
    \cF = \frac{q^2}{2r^4}\,, \quad \cG = 0\,.
\end{align}
This configuration gives the energy-momentum tensor as
\begin{align}
    {T_t}^t = {T_r}^r &= -\cL\,,\\
    {T_\theta}^\theta = {T_\phi}^\phi &= - \cL + \frac{q^2\cL_\cF}{r^4} \,.
\end{align}
Then, under an appropriate scaling of the time coordinate, the Einstein equations can be solved as
\begin{align}
    f = 1 - \frac{2GM}{r} - \frac{8 \pi G}{r} \int^r dr' \, {r'}^2 \cL\left( \frac{q^2}{2 {r'}^4} , 0 \right)\,,
\end{align}
with a constant $M$.

The static configuration should satisfy the equations of motion for the electromagnetic field, from which we can read off a constraint to the Lagrangian,
\begin{align}
    \cL_{\cF \cG}  \left( \frac{q^2}{2 r^4} , 0 \right) = 0\,.
\end{align}
This implies that the Lagrangian does not have linear terms with respect to $\cG$.

\subsection{Electric black holes}
\label{app:electric}

Next, let us find electrically charged solutions.
For this purpose, it is convenient to define the ``conjugate'' field $P_{\mu\nu}$ as\footnote{Here, we defined $\pd F^{\mu\nu} / \pd F^{\alpha \beta} \defi (1/2)(\delta^\mu_\alpha \delta^\nu_\beta - \delta^\nu_\alpha \delta^\mu_\beta)$.} \cite{Breton:2007bza}
\begin{align}
    P_{\mu\nu} &\defi 2\frac{\pd \cL}{\pd F^{\mu\nu} }
    = \cL_\cF F_{\mu\nu} + \cL_\cG \til{F}_{\mu\nu} \,,
    \label{eqA2-1}
\end{align}
and introduce the ``Hamiltonian'' $\cH$ by the Legendre transformation,
\begin{align}
    \cH &\defi \frac{1}{2} P^{\mu\nu} F_{\mu\nu} - \cL
    \notag \\
    &= 2 \cL_\cF \cF + 2 \cL_\cG \cG - \cL\,.
    \label{eqHamil}
\end{align}
Just as $\cL$ is written in terms of two invariants $\cF$ and $\cG$, we can regard $\cH$ as the function of two invariants defined by 
\begin{align}
    \cP \defi \frac{1}{4} P_{\mu\nu} P^{\mu\nu} \,, \quad
    \cQ \defi \frac{1}{4} P_{\mu\nu} \til{P}^{\mu\nu}\,.
\end{align}
Then, we can move from the original $F$ framework, in which one uses the Lagrangian $\cL(\cF, \cG)$ in the analysis, to the so-called $P$ framework, in which $\cH(\cP, \cQ)$ is alternatively utilized.
Note that if the transformation is invertible, which we assume unless otherwise noted, we can recover the $F$ framework from the $P$ framework through
\begin{align}
    F_{\mu\nu} &= 2 \frac{\pd \cH}{\pd P^{\mu\nu}} = \cH_\cP P_{\mu\nu} + \cH_\cQ \til{P}_{\mu\nu}\,,\\
    \cL &= \frac{1}{2} P^{\mu\nu} F_{\mu\nu} - \cH
    \notag \\
    &= 2 \cH_\cP \cP + 2 \cH_\cQ \cQ - \cH\,,
    \label{eqPtoF}
\end{align}
where $\cH_\cP = \pd \cH / \pd \cP$ and $\cH_\cQ = \pd \cH / \pd \cQ$.

Performing the Legendre transformation, the Einstein equations are rewritten in the $P$ framework with the energy-momentum tensor,
\begin{align}
    {T_\mu}^\nu = \cH_\cP P_{\mu\lambda} P^{\nu\lambda} - \delta_\mu^\nu (2\cH_\cP \cP + \cH_\cQ \cQ - \cH)\,.
\end{align}
Similarly, the equations of motion and the Bianchi identity for the electromagnetic field are rewritten as
\begin{align}
    &\nabla_\mu P^{\mu\nu} = 0\,,\\
    &\nabla_\mu \left( \cH_\cP \til{P}^{\mu\nu} - \cH_\cQ P^{\mu\nu} \right) = 0\,,
    \label{eqeleBia}
\end{align}
respectively.

To find static and spherically symmetric solutions with electric charge, we set
\begin{align}
    \frac{1}{2} P_{\mu\nu} dx^\mu \wedge dx^\nu = -\frac{q}{r^2} dt \wedge dr\,.
    \label{eqelebg}
\end{align}
Here, $q$ is a constant corresponding to the electric charge.
On the configurations \eqref{eqssa} and \eqref{eqelebg}, the invariants read
\begin{align}
    \cP = - \frac{q^2}{2r^4} \,, \quad
    \cQ = 0\,,
\end{align}
and the components of the energy-momentum tensor are given by
\begin{align}
    {T_t}^t = {T_r}^r &= \cH\,,\\
    {T_\theta}^\theta = {T_\phi}^\phi &= \cH + \frac{q^2 \cH_\cP}{r^4}\,.
\end{align}
Then, the Einstein equations are solved by
\begin{align}
    f = 1 - \frac{2GM}{r} + \frac{8 \pi G}{r} \int^r dr' \, {r'}^2 \cH\left( - \frac{q^2}{2{r'}^4}, 0 \right)\,,
\end{align}
where $M$ is a constant.

The fact that the electric configuration satisfies the equation \eqref{eqeleBia} gives the constraint to the Hamiltonian as
\begin{align}
    \cH_{\cP\cQ}  \left(- \frac{q^2}{2r^4}, 0 \right) = 0 \,,
\end{align}
from which we can see that the Hamiltonian should not depend on $\cQ$ linearly for the existence of the electric black hole solution.

\subsection{$FP$ duality}
\label{app:FPduality}

From the above analysis, one can read off a duality relation between the $F$ framework and the $P$ framework, which is known as the $FP$ duality.
To see this, let us consider $\til{P}_{\mu\nu}$ as a new electromagnetic field strength $F'_{\mu\nu}$:
\begin{align}
    F'_{\mu\nu} \defi \til{P}_{\mu\nu}\,,
\end{align}
which leads to 
\begin{align}
    \til{F}_{\mu\nu}' &= - P_{\mu\nu}\,,\\
    \cF' &= - \cP\,,\label{eqdual2}\\
    \cG' &= - \cQ\,.\label{eqdual3}
\end{align}
Furthermore, we define a new Lagrangian $\cL'$ by using the original Hamiltonian $\cH$ as
\begin{align}
    \cL'(\cF', \cG') &\defi - \cH(\cP, \cQ)
    \notag \\
    &= - \cH(-\cF', -\cG')\,.
    \label{eqdual}
\end{align}
Consequently, we have
\begin{align}
    \cL'_{\cF'} &= \cH_\cP\,,\\
    \cL'_{\cG'} &= \cH_\cQ\,.
\end{align}
Then, we can see that the equations of motion in the $P$ framework for the original theory defined by $\cL(\cF, \cG)$ are equivalent to those in the $F$ framework for the new theory defined by $\cL'(\cF',\cG')$.\footnote{
    One should note that
    \begin{align}
        \til{P}_{\mu\lambda} \til{P}^{\nu\lambda} = - 2 \cP \delta_\mu^\nu + P_{\mu\lambda} P^{\nu\lambda},
        \notag \\
        \frac{1}{4}\til{P}_{\mu\lambda} \til{P}^{\mu\lambda} = - \cP.
        \notag 
    \end{align}
    }
This relation suggests that the electric solution in the original theory $\cL(\cF, \cG)$ can be obtained as the magnetic solution in the new theory $\cL'(\cF', \cG')$, and vice versa.

Similarly, the analysis of perturbations on magnetic black holes in a given theory can be translated into that on electric black holes in another theory.
In fact, the master equations with translations $\cL \to - \cH$, $\cL_\cF \to \cH_{\cP}$, etc., for potentials \eqref{eqVI11}--\eqref{eqVI22}, \eqref{eqVII11}--\eqref{eqVII22} describe perturbations on the electric black hole in another theory, in which the Lagrangian is given by Eq.\,\eqref{eqPtoF}.

Note that the $FP$ duality relates a magnetic black hole in a given theory to an electric black hole in a different theory in general.
The $FP$ duality is reduced to the symmetry of the theory only if $\cL(\cF, \cG) = \cL'(\cF, \cG) = - \cH(-\cF,-\cG)$. This is true in specific cases, such as the Maxwell electrodynamics $\cL(\cF, \cG) = \cF $ (up to a constant) and Born--Infeld electrodynamics given by Eq.\,\eqref{eqBI} \cite{Gibbons:1995cv}.

\subsection{$FP$ duality in effective field theory}
\label{app:FPdualityEFT}

Here we study the $FP$ duality for effective field theory in nonlinear electrodynamics as an example.
We consider an effective Lagrangian,
\begin{align}
    \cL(\cF, \cG) = \cF - \alpha \cF^2 - \beta \cG^2.
    \label{eqA4-1}
\end{align}
The magnetic black hole solutions in this theory are subjects of the analysis in Sec.\,\ref{sec:EFT} and subsequent sections of this paper.
Here let us construct the Hamiltonian for this effective theory, and obtain the dual theory with a Lagrangian $\cL'(\cF', \cG')$ via Eq.\,\eqref{eqdual}. 
Then, the magnetic black holes in the dual theory describe the electric black holes in the original effective theory.
Note that we will always ignore nonlinear terms with respect to $\alpha$ and $\beta$ in the following descriptions.

First, from Eq.\,\eqref{eqA2-1}, the conjugate field $P_{\mu\nu}$ in the effective theory \eqref{eqA4-1} is given by
\begin{align}
    P_{\mu\nu} = (1-2\alpha \cF) F_{\mu\nu} - 2 \beta \cG \til{F}_{\mu\nu}.
    \label{eqA4-2}
\end{align}
Then we have
\begin{align}
    \cP &\simeq \cF - 4 \alpha \cF^2 - 4 \beta \cG^2,
    \label{eqA4-3} \\
    \cQ &\simeq \cG - 4 \alpha \cF \cG + 4 \beta \cF \cG,
    \label{eqA4-4}
\end{align}
at the linear order in $\alpha$ and $\beta$.
Eqs.\,\eqref{eqA4-3} and \eqref{eqA4-4} can be inverted to obtain 
\begin{align}
    \cF &\simeq \cP + 4 \alpha \cP^2 + 4 \beta \cQ^2,
    \label{eqA4-5}\\
    \cG &\simeq \cQ + 4 \alpha \cP \cQ - 4 \beta \cP \cQ.
    \label{eqA4-6}
\end{align}
From the definition \eqref{eqHamil}, at the linear order in $\alpha$ and $\beta$, the Hamiltonian in the effective theory is given by
\begin{align}
    \cH &= 2 \cL_\cF \cF + 2 \cL_\cG \cG - \cL
    \notag \\
    &= \cF - 3\alpha \cF^2 - 3\beta\cG^2
    \notag \\
    &\simeq \cP + \alpha \cP^2 + \beta \cQ^2
    \label{eqA4-7},
\end{align}
where Eqs.\,\eqref{eqA4-5} and \eqref{eqA4-6} are used in the last line.
Thus, from Eqs.\,\eqref{eqdual2} and \eqref{eqdual}, the Lagrangian of the dual theory to the effective theory \eqref{eqA4-1} is approximately given by
\begin{align}
    \cL'(\cF', \cG') \simeq \cF' - \alpha {\cF'}^2 - \beta {\cG'}^2,
    \label{eqA4-8}
\end{align} 
which coincides with the original effective Lagrangian \eqref{eqA4-1} up to the linear order of $\alpha$ and $\beta$.
This means that, in the effective theory \eqref{eqA4-1}, a black hole with magnetic charge $q$ and a black hole with electric charge $q$ are described by the same geometry, e.g.~Eq.\,\eqref{eqEFTbg}, up to the linear order of $\alpha$ and $\beta$.
This fact is consistent with the analysis of charged black holes in Euler--Heisenberg electrodynamics in Ref.\,\cite{Ruffini:2013hia}.
Furthermore, it is expected that the calculation results for quasinormal frequencies with corrections proportional to $\alpha$ and $\beta$ for magnetic black holes, which are shown in Sec.\,\ref{sec:QNM} in the present paper, are equally valid for electric black holes.

\bibliography{bibliography}

\end{document}